\documentclass{aa}  

\usepackage{graphicx}
\usepackage{txfonts}
\usepackage{caption}
\usepackage{subfig}
\usepackage{subfiles}
\usepackage{dcolumn}
\usepackage{multirow}
\usepackage[normalem]{ulem}
\usepackage{amssymb}
\usepackage{bbding}
\usepackage{color}
\usepackage{booktabs}
\usepackage{comment}
\usepackage{verbatim}

\newcolumntype{d}[1]{D{.}{.}{#1}}

\usepackage{titlesec}

\titleformat{\paragraph}
{\normalfont\normalsize}{\theparagraph}{1em}{}
\titlespacing*{\paragraph}
{0pt}{3.25ex plus 1ex minus .2ex}{1.5ex plus .2ex}

\newcommand{\RRL}[1]{H{#1}$\alpha$}
\newcommand{\methacet}{CH$_3$CCH}
 \newcommand{\hii}{H{\sc ii}}

\usepackage{pifont}

\newcolumntype{d}[1]{D{.}{.}{#1}}
\newcolumntype{.}{D{.}{.}{-1}}
\newcolumntype{;}{D{.}{.}{0}}
\newcommand{\lsun}{L$_\odot$}
\newcommand{\msun}{M$_\odot$}

\begin{document}

   \title{The evolution of young \hii{} regions\thanks{Table \ref{tab:clumps} is available in electronic form
at the CDS via anonymous ftp to cdsarc.u-strasbg.fr (130.79.128.5) or via http://cdsweb.u-strasbg.fr/cgi-bin/qcat?J/A+A/}}

   \subtitle{I. Continuum emission and internal dynamics}

   \author{P.~D. Klaassen\inst{1},
        K.~G. Johnston\inst{2},
        J.~S. Urquhart\inst{3},
        J.~C. Mottram\inst{4},
        T. Peters\inst{5},
        R. Kuiper\inst{6,4},
          H. Beuther\inst{4},
          F.F.S. van der Tak \inst{7,8}
          C. Goddi\inst{9,10}}

   \institute{UK Astronomy Technology Centre, Royal Observatory Edinburgh, Blackford Hill, Edinburgh EH9 3HJ, UK \\ \email{pamela.klaassen@stfc.ac.uk}
   \and
   School of Physics \& Astronomy, E.C. Stoner Building, The University of Leeds, Leeds LS2 9JT, UK
   \and
   Centre for Astrophysics and Planetary Science, University of Kent, Canterbury CT2 7NH, UK
   \and
   Max Planck Institute for Astronomy, K\"onigstuhl 17, 69117, Heidelberg, Germany
   \and
   Max-Planck-Institut f\"ur Astrophysik, Karl-Schwarzschild-Str. 1, D-85748 Garching, Germany
   \and
   Institute of Astronomy and Astrophysics, University of T\"ubingen, Auf der Morgenstelle 10, D-72076 T\"ubingen, Germany
   \and
    SRON Netherlands Institute for Space Research, Landleven 12, 9747 AD, Groningen, The Netherlands
  \and
   Kapteyn Astronomical Institute, University of Groningen, The Netherlands
   \and
   Department of Astrophysics/IMAPP, Radboud University, P.O. Box 9010,
6500 GL Nijmegen, The Netherlands
\and
  ALLEGRO/Leiden Observatory, Leiden University, PO Box 9513, NL-2300 RA Leiden, the Netherlands
   }

\authorrunning{P.D. Klaassen et al}

   \date{Received 07 Aug, 2017; accepted 01 Dec, 2017}

 
  \abstract
   {High-mass stars form in much richer environments than those associated with isolated low-mass stars, and once they reach a certain mass, produce ionised (\hii{}) regions. The formation of these pockets of ionised gas are unique to the formation of high-mass stars (M $>8$ M$_\odot$), and present an excellent opportunity to study the final stages of  accretion, which could include accretion through the \hii{} region itself. }
  {This study of the dynamics of the gas on both sides of these ionisation boundaries in very young \hii{} regions aims to quantify the relationship between the \hii{} regions and their immediate environments.}
   {We present high-resolution  ($\sim$ 0.5$''$) ALMA observations of nine \hii{} regions selected from the Red MSX Source (RMS) survey with compact radio emission and bolometric luminosities greater than 10$^4$ L$_\odot$. We focus on the initial presentation of the data, including initial results from  the radio recombination line \RRL{29}, some complementary molecules, and the 256 GHz continuum emission. }  
   {Of the  six (out of nine) regions with \RRL{29} detections, two appear to have cometary morphologies with velocity gradients across them, and two appear more spherical with velocity gradients suggestive of infalling ionised gas. The remaining two were either observed at low resolution or had signals that were too weak to draw robust conclusions. We also present a description of the interactions between the ionised and molecular gas (as traced by CS (J=5-4)), often (but not always) finding the \hii{} region had cleared its immediate vicinity of molecules.}
   {Of our sample of nine, the observations of the two clusters expected to have the youngest \hii{} regions (from previous radio observations) are suggestive of having infalling motions in the \RRL{29} emission, which could be indicative of late stage accretion onto the stars despite the presence of an \hii{} region. }


   \maketitle
%

\section{Introduction}
\label{sec:intro}

High-mass stars are generally observed in external galaxies because they burn brighter and hotter than their low-mass counterparts. From their light we extrapolate an entire initial mass function worth of stellar mass in those galaxies \citep{KE_ARAA_12}. Massive (proto)stars are expected to provide strong feedback on the formation process with their high UV radiation field.  When the (proto)star becomes massive enough, it begins to heat up, forming an \hii{} region, which ionises the ambient medium, the upper and lower layers of the accretion disk, and sometimes even the outflow \citep[e.g.][]{KW06}. The early evolution of hyper- and ultra-compact \hii{} regions (HC \hii{} and UC \hii,{} respectively)  critically depends on the pre-main-sequence evolution of the embedded massive protostar, including the so-called bloated phase  \citep[e.g.][]{Hosokawa10}, its epoch of disk formation, and accretion rate \citep[e.g.][]{Kuiper13, Kuiper16}. In the bloated phase the massive protostar has a size of 10-100 times its ZAMS radius \citep{Hosokawa10}, which lowers the surface temperature and consequently, the ionising radiation. At the same time, the accretion luminosity and rate are both predicted to be highly variable. These properties  have an important and observable impact on the subsequent evolution of the young \hii{} region \citep[e.g.][]{MKlassen12}. Thus detailed observations of the early evolution of ultra-compact \hii{} regions can provide key constraints on the pre-main-sequence evolution of high-mass stars \citep[e.g.][]{depree14,depree15}.

Forming high-mass stars produce strong stellar winds from an early age,  likely altering the course of their accretion flows in the process \citep[see, for example,][]{Puls08}. When a forming massive protostar begins producing a higher ionising photon flux than can be absorbed by its immediate surroundings,  an \hii{} region forms and expands outwards until the ionisation and recombination balance (i.e. it reaches the Str\"omgren radius).   For most high-mass stars (M$_*\sim10-30$M$_\odot$), accretion will have halted before an \hii{} region can expand outwards \citep{Davies11} because within this mass range, the protostar is likely to bloat as material is accreted, delaying the onset of an \hii{} region \citep[e.g.][]{Hosokawa10,Kuiper13,HP16}.  Beyond this threshold of $\sim30$ M$_\odot$, the star can contract back to a main-sequence configuration even for high rates of ongoing accretion. Stars with masses much greater than this (e.g. $> 60$ M$_\odot$) exist in our Galaxy, therefore something must allow for accretion onto the forming high-mass star in the presence of an \hii{} region.  

Very little is known about how that accretion continues, thus, further observational constraints are required.  Such observations should include the dynamics of both the ionised and molecular gas components of very young (i.e. HC-) \hii{} regions. How is the ionised gas moving? Is it contracting or expanding? Is it rotating and or infalling? How is the molecular gas moving in comparison?  Could it be feeding (i.e. providing a collapsing mass reservoir) the ionised gas component?  These types of questions can only be answered using comprehensive (i.e. matched sensitivity and resolution) observations of a large number of these young \hii{} regions.

Here we present the first results of such a survey, including the largest sample of high-resolution observations of small (i.e. hyper-compact, < 0.01 pc)  \hii{} regions observed so far, which can be used to quantify the dynamics of both the ionised and molecular gas.   From this coherent dataset we can begin to draw conclusions about the nature of the interaction between \hii{} regions and their environments.  

Our sample is drawn from \hii{} regions in the southern sky with known submm continuum fluxes and high-resolution radio (5GHz) continuum images that constrain the size of the \hii{} region. With our new ALMA observations we can trace the dynamics of the ionised gas to quantify its expansion into the ambient cloud. By understanding the dynamics of the molecular gas into which it is expanding, we determine what effects the \hii{} region has on its environment, whether the original infall/rotation of the gas is disrupted, and how much energy the \hii{} region feeds back into its surroundings.

Below, in Section \ref{sec:obs}, we present our source selection criteria and briefly outline our data reduction methods.  In Section \ref{sec:results} we present our analysis of the clustering and the masses of the cores in each region, and the dynamics of both the ionised and molecular gas species emitting on small scales.  We then compare the observed ionised gas velocity structures to those predicted in models.  We summarise our findings in Section \ref{sec:conclusions}.  Additional details concerning the combination of the multi-configuration ALMA data can be found in Appendix \ref{sec:short_spacing} and in Appendix \ref{subsec:molec_removal} we give a detailed description of the \texttt{XCLASS} modelling required to separate the overlapping ionised and molecular emission in our \RRL{29} spectral window.

\section{Observations}
\label{sec:obs}

The Red MSX Source (RMS) Survey has identified $\sim$600 \hii{} regions distributed throughout the Galactic plane \citep{RMS2013}. This sample of \hii{} regions is the largest and most well characterised yet compiled, providing reliable distances \citep{Urquhart13}, bolometric luminosities derived from spectral energy distribution (SED) modelling \citep{Mottram11b}, and targeted high-resolution ($\sim1''$) radio continuum observations made with the ATCA and VLA \citep{Urquhart07,Urquhart09}.

From this sample we selected \hii{} regions that are located in the southern Galactic plane, are  associated with compact radio emission (radius $< 5''$), have bolometric luminosities greater than 10$^4$ L$_\odot$, and are located between 3 and 5.5 kpc to keep the linear scales roughly consistent. We  used the ATLASGAL submm emission maps to refine our sample by selecting sources associated with massive and small dense clumps \citep[$<35''$, ][]{Urquhart14}.

Our selection criteria provide a distance limited sample of nine luminous HC and UC \hii{} regions that are still embedded in their natal environment. In Table \ref{tab:obs_params} we show the positions, distances to, and luminosities of our sample.  These sources are ideally located for ALMA and the compactness of these sources means they can be observed with a single pointing with ALMA Band 6 observations ($\sim$250 GHz, or a wavelength of ~1.2 mm).

\begin{table*}
\begin{center}
\caption{Source locations and beam properties}
\begin{tabular}{ccrc|ccrc}
\hline \hline
\multicolumn{4}{c|}{Source properties} & \multicolumn{4}{c}{Observation properties}\\
\hline
 Name & \multicolumn{2}{c}{Pointing centre} & V$_{\rm LSR}$& \multicolumn{3}{c}{Synthesised beam} & Continuum\\
  & RA & DEC &&  Bmaj & Bmin & BPA & rms noise \\
  & (h:m:s) & (d:m:s) &(km s$^{-1}$)& ($''$) & ($''$) & ($^\circ$) & mJy/beam\\ 
  \hline

\multicolumn{8}{c}{Targets near G305  (3 regions)}\\

\hline
G302.02+00.25 & 12:43:31.52 & $-$62:36:13.6 &  $-$37.1& 0.61 & 0.58 & 28.38 & 0.15 \\
G302.48$-$00.03 & 12:47:31.75 & $-$62:53:59.4 &   $-$37.1& 0.61 & 0.57 & $-$174.51 & 0.17 \\
G309.89+00.40 & 13:50:35.54 & $-$61:40:21.4 &   $-$57.6&0.69 &  0.62 & 6.17 & 0.15 \\
\hline

\multicolumn{8}{c}{Targets near G335 (6 regions)}\\

\hline
G330.28+00.49 & 16:03:43.28 & $-$51:51:45.6 &  $-$93.5& 0.71 & 0.53 & 87.19 & 0.13\\
G332.77$-$00.01 & 16:17:31.13 & $-$50:32:35.7& $-$95.6& 0.62 & 0.46 & 87.62 & 0.18 \\
G336.98$-$00.18 & 16:36:12.42 & $-$47:37:57.5  &$-$75.1& 0.63 & 0.47 & $-$90.00 & 0.13 \\
G337.63$-$00.08 & 16:38:19.02 & $-$47:4:50.7 & $-$56.5&0.61 & 0.46 & 90.00 & 0.12 \\
G337.84$-$00.37 & 16:40:26.68 & $-$47:7:13.1  & $-$40.4&0.62 & 0.47 & $-$89.07 & 0.18 \\
G339.11+00.15 & 16:42:59.58 & $-$45:49:43.6  & $-$78.2&1.68 & 1.11 & $-$84.79 & 0.51 \\
\hline \hline
\label{tab:obs_params}
\end{tabular}
\end{center}
\end{table*}

In Table\,\ref{tbl:hii_parameters} we present the radio parameters of the \hii{} regions and their host cores. The observing pointing centres are taken from the position of the peak radio flux densities and the distances with their associated uncertainties are computed using the \citet{reid2016} rotation model. The masses and bolometric luminosities are recalculated from the values given in \citet{Urquhart14} using the updated source distance and assuming an isothermal dust temperature of 30\,K on single dish scales. Allowing for a variation in the temperature of $\pm$5\,K and taking account of distance uncertainties, we estimate a measurement uncertainty in the mass of $\sim$30-50\,per\,cent, while other systematics could be pushing the formal uncertainties to a factor of 2-3.

\begin{table*}
\begin{center}\caption{Summary of radio properties of the \hii\,regions and their host molecular clumps. The luminosity in Column 7 is that of the \hii{} region, and the final column gives the radio derived spectral index. For the targets which were not detected in all of the ATCA and VLA wavebands in \citet{Urquhart07,Urquhart09}, no spectral index ($\alpha$) was determined. }
\label{tbl:hii_parameters}
\begin{minipage}{\linewidth}
\small
\begin{tabular}{lcc.@{$\pm$}l.....}
\hline \hline
\multicolumn{1}{c}{Source Name}& \multicolumn{1}{c}{RA} & \multicolumn{1}{c}{Dec.} & \multicolumn{2}{c}{Distance} & \multicolumn{1}{c}{Log[M$_{\rm{clump}}$]} & \multicolumn{1}{c}{Log[L$_{\rm{Bol}}$]} & \multicolumn{1}{c}{Radio Size} & \multicolumn{1}{c}{Flux\tablefootmark{(a)}$_{\rm{6\,cm}}$}& \multirow{2}{*}{$\alpha$}\\
\multicolumn{1}{c}{}& \multicolumn{1}{c}{(J2000)} & \multicolumn{1}{c}{(J2000)} & \multicolumn{2}{c}{(kpc)}  & \multicolumn{1}{c}{(\msun)} & \multicolumn{1}{c}{(\lsun)} & \multicolumn{1}{c}{(\arcsec)} & \multicolumn{1}{c}{(mJy)} & \\
\hline
G302.02+00.25 	&	 12:43:31.49 	&	 $-$62:36:13.7 	&	4.26	&	0.98	&	2.76	&	4.16	&	2.24	&	59.5	&	-0.03	\\
G302.49$-$00.03 &	 12:47:31.76 	&	 $-$62:53:59.6 	&	3.39	&	0.40	&	2.61	&	3.75	&	2.07	&	23.4	&	0.03	\\
G309.89+00.40 	&	 13:50:35.54 	&	 $-$61:40:21.4 	&	5.41	&	1.27	&	3.12	&	4.40	&	1.79	&	1.2	&	\multicolumn{1}{c}{$\cdots$}	\\
G330.28+00.49 	&	 16:03:43.26 	&	 $-$51:51:45.9 	&	5.45	&	0.56	&	2.95	&	4.13	&	2.00	&	44.9	&	-0.08	\\
G332.77$-$00.01 &	 16:17:31.13 	&	 $-$50:32:35.7 	&	5.67	&	0.57	&	3.22	&	4.16	&	1.14	&	24.6	&	\multicolumn{1}{c}{$\cdots$}	\\
G337.63$-$00.08 &	 16:38:19.02 	&	 $-$47:04:51.0 	&	3.82	&	0.43	&	2.88	&	4.01	&	2.20	&	31	&	0.01	\\
G337.84$-$00.37 &	 16:40:26.68 	&	 $-$47:07:13.1 	&	2.98	&	0.47	&	2.55	&	4.59	&	3.32	&	11.1	&	0.63	\\
G336.98$-$00.18 &	 16:36:12.43 	&	 $-$47:37:58.0 	&	4.67	&	0.41	&	2.51	&	4.43	&	1.47	&	18	&	0.52	\\
G339.11+00.15 	&	 16:42:59.58 	&	 $-$45:49:43.6 	&	4.96	&	0.40	&	2.93	&	4.24	&	1.98	&	42.9	&	-0.04	\\
\hline
\end{tabular}\\
\tablefoottext{a}{Integrated Flux at 6 cm}
\end{minipage}
\end{center}
\end{table*}

The ALMA data presented here (from project 2013.1.00327.S, PI: P. Klaassen) consists of two groupings of targets based on co-observability on the sky (groups of \hii{} regions within 10$^\circ$ of each other).  Each group was observed  with the Atacama Compact Array (ACA) and in two configurations with the 12 m baseline array (one compact, one extended) covering baseline lengths from 15m to $\sim$784 m.  The integration times, calibrators (and their purposes), observing date, and  minimum and maximum baseline lengths used in each of the executions contained in this set of observations are listed in Table \ref{tab:calibrators}. The spectral set-up is given in Table \ref{tab:spec_setup}.   G339.11+00.15 was not observed in the most extended configuration. Hereafter, when referring to the targets of this study, we refer to such target with a shortened name consisting of the targets Galactic longitude (e.g. G302.02$+$00.25 becomes G302.02).

\begin{table}
\caption{Spectral Setup of the observed regions. The primary targetted species of each spectral window are listed, with the note that spectral windows 1 and 2 were combined to get the full CH$_3$CN (J=12-11) k-ladder.}
\begin{tabular}{cd{3.2}ccd{3.1}d{1.3}}
\hline\hline
SPW & \multicolumn{1}{c}{Freq} & Species & \multicolumn{1}{c}{N. Chan}
 & \multicolumn{1}{c}{Vel. BW} & \multicolumn{1}{c}{Vel. Res.}\\
 
& \multicolumn{1}{c}{(GHz)} &  &
 & \multicolumn{1}{c}{(km/s)} & \multicolumn{1}{c}{(km/s)}\\
\hline
1	&257.403	&CH$_3$CN	&3840		&545.8	&0.284\\
2	&257.127	&CH$_3$CN	&1920		&273.2	&0.285\\
3	&256.302	&H29$\alpha$	&1920		&274.1	&0.258\\
4	&244.936	&CS	&1920		&286.8	&0.299\\
5	&242.998	&SO$_2$	&1920		&289.1	&0.301\\
6	&241.616	&SO$_2$	&960		&145.4	&0.303\\
\hline
\end{tabular}
\label{tab:spec_setup}
\end{table}

To ensure proper weighting of the different configurations when combining datasets, all of the data were manually reduced (i.e. not using the ALMA pipeline) in CASA 4.2.2 \citep{CASA}. For each execution block, calibrations were made for water vapour, system temperatures, and antenna positions followed by bandpass, phase, and amplitude (flux) calibration.   In the instances of multiple executions (e.g. ACA observations) an additional level of flux calibration was conducted to ensure consistent fluxes across each execution.

For each target and in each configuration, the continuum emission was strong enough for self-calibration, which was done in phase only. Additional amplitude self-calibration was found to not improve the signal-to-noise ratio, and was therefore omitted.  The only exception was G332.77, for which self-calibration was not possible (peak S/N = 13) in the highest resolution dataset.

The continuum emission was derived from the line free portions of each spectral window and then split out for each science target (using \verb+uvcontsub+), and used for self-calibration of the data.    These self-calibration solutions were then applied to the line data.

\section{Results}
\label{sec:results}

We present analyses of some of the data observed with ALMA in this paper. A full analysis of all the  data obtained for this project and their implications are presented in follow-up papers. In this work, we focus on the continuum emission, ionised gas observed within the \hii{} regions, and some of the small scale molecular emission (in CS and \methacet{}) surrounding the \hii{} regions to put the analysis of the ionised gas into context.  Full analysis of the molecular environments and the chemistry of these regions is left to future work (e.g. Klaassen et al. in prep, van der Tak et al. in prep).

The analysis is broken down into a number of subsections; continuum (Section \ref{subsec:continuum}), ionised and molecular gas components (Sections \ref{subsec:ionised}, \ref{sec:models} and \ref{subsec:molecular}) are discussed here. Because the ionised and molecular spectra (primarily the \RRL{29} and \methacet) were blended, there is an additional section detailing how their emission components were disentangled that is presented in Appendix \ref{subsec:molec_removal}. 

Figures \ref{fig:G302.02} through \ref{fig:G332.77} show the 256 GHz continuum emission and bulk molecular gas emission (as traced by CS) for each of the regions studied here. Additionally, for those sources for which \RRL{29} was detected (Figures \ref{fig:G302.02} - \ref{fig:G339.11}), we show both maps and spectra of this emission.

\subsection{Continuum and cores}
\label{subsec:continuum}

\subsubsection{Core images}

In the top left panels of Figures \ref{fig:G302.02} through \ref{fig:G332.77}, we present the 256 GHz continuum emission for each of our observed \hii{} regions.  Overplotted in  red contours is the 5 GHz emission (see Section \ref{sec:clumps}) and the scale bar in the top left corner represents the thermal Jeans Length in each region (see Section \ref{sec:Jeans_length}).  The cores identified in each region are shown with yellow contours and the number of each core is labelled in white from brightest to dimmest. These are numbered in the same order as presented in Table \ref{tab:clumps}, which presents the derived core properties in each region.

\subsubsection{Radio continuum and SED analysis}

The radio emission is relatively compact, is typically of a few arc-seconds in size, and has fluxes of up to $\sim$ 60 mJy at 5 GHz (see Table \ref{tab:clumps}).  All but two of the targets have both 5 and 8.6 GHz detections with ATCA. For those detected at both frequencies, spectral indices ($\alpha$) were calculated and generally found to be of order zero, as expected for optically thin thermal emission.  There are two exceptions (G336.98 an G337.84), whose spectral indices are closer to one (see Table \ref{tbl:hii_parameters}), and are indicative of optically thin thermal emission \citep[e.g.][]{Zapata08} or ionised jets \citep[e.g.][]{Purser16}. In the analysis presented here, we favour the optically thin scenario over the ionised jet because of the relationship between the  8 GHz and IR luminosities of these targets. When plotted against each other \citep[as in Figure 6 of ][]{HoareFranco07}, these targets, with L$_{8GHz}$ $>$ 10$^{13}$ W Hz$^{-1}$ and L$_{bol}$ $>$ 10$^4$ L$_\odot$, lie in the HC \hii{} region regime, more than one dex in L$_{8GHz}$ above the jet sources.    These are likely the youngest \hii{} regions in our sample and are also the regions directly associated with the masers plotted in Figures \ref{fig:G302.02}, \ref{fig:G330.28}, \ref{fig:G336.98}, \ref{fig:G337.63}, and \ref{fig:G337.84}.

Using the radio spectral indices, we quantified the amount of free-free emission expected in our 256 GHz continuum observations. Subtracting that value from the observed emission results in the thermal dust component of the emission from each region (see Table \ref{tab:ff-cont}). In most cases, the free-free component represents more than 50\% of the 256 GHz continuum flux for the mm  core associated with the \hii{} region (see the core analysis below in Section \ref{sec:clumps}).

\subsubsection{Core  boundaries and masses}
\label{sec:clumps}

Each of the systems observed in this sample appears to have at least moderately ($>$2) clumpy structures in the continuum. In quantifying that so-called clumpiness, we can determine properties about the cluster, including its structure and the mass of each core.  To do this, we used the \texttt{Fellwalker} package \citep{fellwalker}, which uses a watershed algorithm to determine the edges of the boundaries between cores in the input continuum image. This watershed algorithm finds the steepest downwards gradients in emission from each peak (like water running down a hill) to find where it pools (at the noise threshold). The final multi-configuration continuum images were clipped at 3$\sigma$ (with $\sigma$ given in Table \ref{tab:obs_params}), and in each case, the algorithm found at least two cores in each image.  The cores themselves are shown with yellow contours and are numbered in white in the top left panels of Figures \ref{fig:G302.02} to \ref{fig:G332.77}. The properties of the cores associated with the \hii{} regions are shown in Table \ref{tab:ff-cont} and the full listing of core properties is presented in Table \ref{tab:clumps}.

The outputs of \texttt{Fellwalker} include the positions of the peaks of each core, the enclosed area (in square arcsec), and summed intensity within the core.  The resultant intensities are listed in  Table \ref{tab:clumps}.  From the flux of each core, and certain assumptions about the targets (such as filled beams and ambient temperatures), we derived core masses as described below.

The mass of each core is derived under the assumption that the emission is dominated by the thermal dust continuum, using equation 6 of \citet{Hildebrand83} such that

\begin{equation}
 M_c = \left(\frac{F_{\nu}D^2}{B(\nu,T)}\right)\left(\frac{4}{3}\frac{a}{Q{\nu}}\right)
\rho\left(\frac{M_g}{M_d}\right)
\label{eqn:dust_mass}
,\end{equation}

\noindent where $F_{\nu}$ is the observed dust flux density, $D$ is the distance to source, $B(\nu,T)$ is the Planck blackbody dust intensity, $a$ is  the assumed dust grain size, $Q_\nu$ is the dust emissivity, and ${M_g}/{M_d}$ is the gas to dust mass ratio \citep[assumed to be 100; e.g.][]{Draine04}.

 In Table \ref{tab:clumps} we show the results of these calculations at three temperatures (20, 50, and 100 K) to show how the masses change with assumed temperature. There is only potential for free-free emission contamination in the mass estimates for (at most) one core in each cluster.  The free-free corrected values are presented in Table \ref{tab:ff_corrected_continuum} for the contaminated cores.  

We additionally use equation \ref{eqn:dust_mass} to quantify our point source mass sensitivity in a single beam with the rms noise limits listed in Table \ref{tab:obs_params} to find that we are sensitive to masses and column densities  at the 10$^{-2}$ M$_\odot$, or 3.6$\times10^{21}$ cm$^{-2}$ limits, respectively.

\subsubsection{Core separation and Jeans lengths}
\label{sec:Jeans_length}

The clumpy nature of the continuum emission allowed us to analyse the fragmentation of each region on scales of the Jeans lengths \citep[see e.g.][]{Wang14}. While we did not have enough cores to create a significant core mass function, we can compare the observed fragmentation scales to the Jeans length in each region.

The Jeans lengths were calculated using
\begin{eqnarray}
\lambda_J &=& \left(\frac{\pi c_s^2}{G\rho}\right)^{1/2}\\
\lambda_J &=& 0.4 {\rm pc} * \left(\frac{c_s}{0.2 {\rm \,km\, s}^{-2}}\right)  \left(\frac{10^3 {\rm \,cm}^3}{n}\right)^{1/2}
\label{eqn:jeans}
,\end{eqnarray}

\noindent where $c_s$ is the sound speed, $\rho$ is the average mass density in each region, and $n$ is the volume number density of each region.  The sound speed can be calculated from the temperature of each region as $c_s = \sqrt{(kT/m)}$, where $k$ is the Boltzmann constant and $m$ is the mean molecular mass of the gas \citep[set here to 2.8*$m_H$,][]{Kaufmann08}.

The sound speed and volume density of the molecular gas were calculated from the \texttt{XCLASS} analysis of the \methacet{} presented in Section \ref{sec:remove_molecules}, and roughly correspond to 0.5 km s$^{-1}$ (T$\sim$ 70-100 K) and $\sim10^7$ cm$^{-3}$ (from the column density and the assumption of spherical symmetry). This gives thermal Jeans lengths of $\sim2000$ au in each region, which is a lower limit because the volume density of the progenitor cores was lower. The larger scale densities, from which these cores fragmented, are much closer to 10$^{5}$ cm$^{-3}$ \citep[see, for instance,][]{Csengeri16}, resulting in a Jeans length 10 times larger, and more consistent with the fragmentation lengths seen in these regions. These values are presented in Table \ref{tab:clump_sep}. The comparison can also be made to the turbulent Jeans length \citep[see e.g.][]{Wang14}. In this instance, the turbulent line width (standard deviation of the line width, not the FWHM) replaces the thermal line width (sound speed) in Equation \ref{eqn:jeans}.  This turbulent Jeans length corresponds to $\sim$ 0.3 pc using large scale densities.

With the results of the \texttt{Fellwalker} analysis presented above, we can quantify the separation between the cores using a minimum spanning tree method \citep[e.g.][]{Allison09}.  Using the core peak coordinates listed in Table \ref{tab:clumps}, we determined the spacing between cores using the algorithms available in \verb+astroML+ \citep{astroML}.  The mean (and standard deviation, $\sigma$) of these separations are given in Table \ref{tab:clump_sep}, and these separations can then be compared to the Jeans length for each region, as given in Table \ref{tab:clump_sep}.

 In all cases except G339.11, for which we have lower resolution data, we resolved the thermal Jeans length. Of the eight regions with high enough resolution observations, all have mean (and indeed minimum) core separations greater than the thermal Jeans length.  However half of the regions have minimum core separations similar to the turbulent Jeans length  (see Table \ref{tab:clump_sep}), which suggests that turbulence was a dominant factor in determining the fragmentation in these regions.

\begin{table*}
\begin{center}
\caption{Derived Jeans lengths (given in au), and core separation statistics, including the mean    and standard deviation ($\sigma$) of the core separations.     Generally, the Jeans length (L$_{\rm Jeans}$) is resolved in these observations. The Turbulent Jeans Length,     for comparison, is of order 60000.0 AU.}
\begin{tabular}{cccccccc}
\hline \hline
Target & No. clumps & L$_{\rm Jeans}$ & Mean Sep. & $\sigma$ & Min Sep. & L$_{Jeans}$/Mean Sep. & Max Sep. \\
\hline\hline
G302.02 & 13 & 18740 & 17900 & 6500 & 5000 & 1.05 & 27400 \\
G302.48 & 11 & 18740 & 12900 & 6800 & 5700 & 1.45 & 28200 \\
G309.89 & 12 & 18740 & 22600 & 9900 & 12500 & 0.83 & 40300 \\
G330.28 & 7 & 20772 & 27400 & 13800 & 10800 & 0.76 & 50900 \\
G332.77 & 2 & 18740 & 38700 & 0 & 38700 & 0.48 & 38700 \\
G336.98 & 4 & 12867 & 11200 & 5300 & 7000 & 1.15 & 18600 \\
G337.63 & 7 & 22286 & 9100 & 5000 & 3000 & 2.45 & 17600 \\
G337.84 & 9 & 18740 & 13400 & 7500 & 4600 & 1.40 & 25400 \\
G339.11 & 5 & 15838 & 27200 & 10900 & 15100 & 0.58 & 40100 \\
\hline
\end{tabular}
\label{tab:clump_sep}\end{center}
\end{table*}

\begin{table}
\caption{Radio (5 GHz) flux of each \hii{} region and spectral  $\alpha$ (updated from \citet{Urquhart14}). The total flux from the \hii{} region bearing \texttt{Fellwalker} core (S$_{\rm tot}$) is subsequently listed as a reminder from Table \ref{tab:clumps} because the free-free extrapolated flux (S$_{\rm FF}$) is then removed from the 256 GHz continuum flux to give the expected contribution to the continuum emission (S$_{\rm dust}$). All fluxes are listed in units of mJy. Using the dust mass assumptions described in Section \ref{sec:clumps} and a temperature of 70 K, the resultant dust masses are presented in units of M$_\odot$.}
\label{tab:ff_corrected_continuum}
\begin{tabular}{lrrrrrr}
\toprule
  Source &  S$_{\rm 5 GHz}$ &   $\alpha$ &  S$_{\rm tot}$ &  S$_{\rm FF}$ &  S$_{\rm dust}$ &  M$_{\rm dust}$ \\
\midrule
 G302.02 &             59.5 &       -0.03 &           62.1 &          53.2 &             8.9 &            29.6 \\
 G302.48 &             23.4 &        0.03 &           93.9 &          26.2 &            67.7 &           142.4 \\
 G309.89 &              1.2 &        \ldots &           36.3 &           \ldots &             \ldots &             \ldots \\
 G330.28 &             44.9 &       -0.08 &           79.9 &          33.3 &            46.6 &           253.5 \\
 G332.77 &             24.6 &        \ldots &            3.5 &           \ldots &             \ldots &             \ldots \\
 G336.98 &             18.0 &        0.52 &          243.9 &         126.7 &           117.1 &           312.8 \\
 G337.63 &             31.0 &        0.01 &           53.1 &          32.2 &            20.9 &            33.9 \\
 G337.84 &             11.1 &        0.63 &          135.3 &         118.1 &            17.2 &            68.5 \\
 G339.11 &             42.9 &       -0.04 &          189.3 &          36.9 &           152.4 &           686.0 \\
\bottomrule
\end{tabular}
\label{tab:ff-cont}

\end{table}

\subsection{Ionised gas}
\label{subsec:ionised}

Often pressure broadening dominates the line widths of Radio Recombination Lines (RRLs) at longer wavelengths. This pressure broadening comes primarily from the electrons in dense \hii{} regions \citep[cf.][]{BS72,Brown78}, which for a $n_e$ = 10$^6$ cm$^{-3}$  \hii{} region, can result in line widths greater than 400 km s$^{-1}$ (for H$n\alpha$, where $n$ = 100, at $\sim$6.5 GHz). However, in our case (with $n$ = 29), the pressure broadening contribution for such an \hii{} region is only a few m s$^{-1}$.  Pressure broadening, which scales linearly with electron density, manifests as a Lorentzian line profile, and not a Gaussian line profile as many other broadening mechanisms do \citep[see e.g.][]{KZK08,Roberto12}.  The result of the blend of  Gaussian and Lorentzian profile is a Voigt profile.  For our \RRL{29} observations we performed both Gaussian and Voigt profile fitting and find our data are best fit by Gaussian profiles alone (or Voigt profiles with no Lorentzian component). From these pressure broadening non-detections, we can place constraints on a number of \hii{} region properties, including temperature and electron density.  For these constraints, we use the larger of the uncertainty on the Gaussian width of the \RRL{29} line and the velocity resolution of our observations (0.25 km s$^{-1}$).  

The \RRL{29} spectra (averaged over the emission area), best fit Gaussian profiles, and resulting residuals are all presented in the bottom left panels of Figures \ref{fig:G302.02} to \ref{fig:G339.11}, with the Gaussian parameters presented in Table \ref{tab:RRL_gaussians}, where the expected molecular V$_{\rm lsr}$ is also given for reference.

It appears that for each region with an \RRL{29} detection, the RRL emission is offset from the reported source V$_{\rm LSR}$ (e.g. those given in Table \ref{tab:RRL_gaussians}) by a few km s$^{-1}$, suggesting either that the source velocity is uncertain by a few km s$^{-1}$, or that the \hii{} region is moving at a slightly different velocity to its surroundings.   It is likely that there is a mixture of these two cases, as the CS emission shown for the same areas in Figures \ref{fig:G302.02} - \ref{fig:G332.77} are a mixture of having the same offsets (e.g. G302.02), which are centred at the same velocities but with smaller velocity ranges (e.g. G337.63) and are consistent with the LSR velocities previously known for the source and offset from the RRL emission (e.g. G337.84).

Below we describe the detection rate of \RRL{29} (Section \ref{sec:H29a_detection_rate}),  properties of the \hii{} regions we derived from the \RRL{29} emission (Section \ref{sec:HII_props}), and our analysis of the velocity gradients seen in each region (Section \ref{sec:velo_grads}), and their implications.

\subsubsection{RRL detection rate}
\label{sec:H29a_detection_rate}

After the fitting and removal of molecular contamination, \RRL{29} was detected in a total of six out of nine of our \hii{} regions.   The top right panels of Figures \ref{fig:G302.02} through \ref{fig:G339.11} show the integrated intensity (moment zero) and intensity weighted velocity (moment one) maps of the \RRL{29} emission in each of the six regions, where regions that have \methacet{} subtraction applied are noted with asterisks. In all but one case (G337.63), the \RRL{29} emission is at or near the position of the brightest continuum source.  For G337.63, the \RRL{29} emission is detected towards the fourth brightest continuum peak (labelled as `4' in Figure \ref{fig:G337.63}).  We note that for G330.28, the \RRL{29} is slightly offset from the continuum peak, however it is still localised to the brightest core within the observed field of view. 

There are a number of reasons why the \RRL{29} was not detected in three of our targets. For G332.77, our signal-to-noise ratios are generally (e.g. in the continuum and molecular emission) very low, suggesting any RRL emission would be below our detection threshold.   This is not the case for G302.48 and G309.89, for which the continuum and molecular emission are well detected. This then suggests that the nature of these sources is what leads to the non-detections.  The principle physical reasons are related to potentially low temperatures, \hii{} regions that are too diffuse (i.e. old) for the detection of the high density lines, or that the star(s) powering the \hii{} regions are not (yet) massive enough to produce high frequency RRLs. G309.89 for instance, has been detected in \RRL{109} and \RRL{110} \citep{CH87}, but remains undetected in \RRL{29} in our sample. Using the first equation in Section 3 of \citet{CH87}, their derived temperature of 6400 K for this region, and our 256 GHz continuum peak for the core (36 mJy, see Table \ref{tab:clumps}) as an upper limit to the unknown fraction of the continuum that comes from free-free emission, we find an expected \RRL{29} line peak below our line sensitivity threshold.

\subsubsection{Derived \hii{} region properties}
\label{sec:HII_props}

If we assume that the Gaussian line width is purely thermal, we can derive an upper limit to the electron temperature required to produce the observed line width.  These values are presented in Table \ref{tab:HII_prop}. For $n=29$, the natural line width is of order 1 m s$^{-1}$ and therefore  is not a significant contribution to the line width \citep[see e.g. Chapter 2 of ][]{Gordon02}.

With an upper limit to the pressure broadening coming from the uncertainties on the Gaussian fits to the \RRL{29} emission, an upper limit on the electron density is derived via \citep{BS72}

\begin{equation}
\frac{\Delta\nu_I}{\nu_0} = 1.43\times10^{-5}\left(\frac{n}{100}\right)^{7.4}\left(\frac{10^4 \\ {\rm K}}{T}\right)^{0.1}\left(\frac{n_e}{10^4 \\ {\rm cm}^3}\right)
,\end{equation}

\noindent where $\Delta\nu_I$ is the upper limit to the pressure broadened width of the line, $n$ is the quantum number of the transition ($n=29$), $T$ is the electron temperature (in K) derived above, and $n_e$ is the electron density (in cm$^{-3}$). Through this equation, we see that the lack of pressure broadening constrains the electron densities in these regions to be less than $5\times10^7$ cm$^{-3}$. The electron density  requirement for a region to be classed as hyper compact is of order 10$^{6}$ cm$^{-3} $\citep{Hoare07,Kurtz05}, which is lower than the limits suggested by our observations, and it is therefore conceivable that these are indeed still hyper-compact \hii{} regions.  This is consistent with the sizes of the  \RRL{29} emitting regions as they are all smaller than the 10,000 au approximate upper limit on the size of an HC\hii{} region \citep[see for instance][]{Hoare07}.

With the electron density upper limit in hand, we can further (loosely) extrapolate to upper limits on the mass of the gas in each \hii{} region (but not the mass of the ionising source).  From the area enclosed in the moment zero maps presented in Figures \ref{fig:G302.02} - \ref{fig:G339.11} and the assumption of spherical symmetry, we can approximate a volume for each \hii{} region.  Multiplying that volume by the electron density (which should be the same as the proton density in the \hii{} region), we determine the number of free protons in the \hii{} region. This can then be translated to the \hii{} region masses presented in Table \ref{tab:HII_prop}.  We find that these limits are very high, which suggests that the upper limits on the electron densities (especially in the case of G339.11) are not very strict (i.e. that the electron densities could be an order of magnitude lower).

An estimate of the free-free continuum emission can be quantified from the integrated line flux and temperature \citep{Brown78} as follows:

\begin{equation}
\frac{\int T_L dv}{T_c} \sim 6.76\times 10^3 \nu^{-1.1}T_e^{-1.15}
,\end{equation}

\noindent where the integrated line flux ($\int T_L dv$) is taken from the Gaussian fits presented in Table \ref{tab:RRL_gaussians} and is transformed  into  units of mJy using the synthesised beam sizes presented in Table \ref{tab:obs_params}. 

The free-free continuum emission levels derived using this method are anywhere between five and 40 times lower than those derived from extrapolating the radio spectral index to the mm.  It should be noted that this equation assumes optically thin emission, which, as we see in Section \ref{sec:models}, may  not be the case for at least some of our \hii{} regions.

\subsubsection{RRL velocity gradients}
\label{sec:velo_grads}

As can be seen in the first moment maps of  \RRL{29} (Figures \ref{fig:G302.02} through \ref{fig:G332.77}), none of our \hii{} regions have the linear velocity gradients usually associated with rotation.  The resolution achieved for G339.11 is much coarser than that for the others (of order 1.4$''$ instead of $\sim0.5''$) so its dynamical/morphological type is less clearly defined.   Its appearance (see Figure \ref{fig:G339.11}) does not suggest it to be cometary, and whether  it has the same dynamical signature of infall on the large size scales probed by our observations is unclear. Higher resolution observations would be required to distinguish this from multiple cores with our beam.

Our  \RRL{29} detection in G337.84 is a weak detection and therefore we cannot properly constrain a velocity gradient. For G330.28 and G337.63, the first moment maps of the RRL emission are more indicative of a constant velocity bow/arc with a blue or red (respectively) tail.   In these cases, indeed, the integrated intensity maps of the RRL emission are also the most cometary in appearance.  These velocity structures are similar to those seen by \citet{Immer14} in \RRL{66} in DR21 on much larger scales (in a much larger and older \hii{} region), which they interpret as being due to bow shocks.

It is also in these regions that we see the CS emission wrapping around the \hii{} region (see Figures \ref{fig:G330.28} and \ref{fig:G337.63}), with a dearth of emission `downwind' from the \hii{} region.  In general, the peak of the CS emission is offset from the peak of the \RRL{29} emission. However it is in these two regions (G330.28 and G337.63) that the most striking offsets occur: the CS appears to be upwind (i.e. closer to the bow than the tail) of the \hii{} region and little to no emission overlaps with the \hii{} region itself. If the \hii{} region is pushing through the ambient material, this offset is likely due to the molecular gas having been compressed at the head of the cometary region, which leaves a cavity in its wake. If the offset is caused by expansion of the \hii{} region, then it is a product of the density gradient, where one edge of the region pushes against (and compressing) the dense material while the other edge is able to break out of the ambient gas and push out the less dense material as it expands.

 For two of our targets, (G302.02 and G336.98, and potentially G339.11, but at lower resolution), the RRL shows a concentric velocity gradient, the so-called `bulls-eye' morphology \citep[see, for instance,][]{Sollins05}.  Similar (e.g. redshifted towards the centre) profiles were predicted for optically thick RRL emission in infalling \hii{} regions by \citet[][ see Section \ref{sec:models}]{Peters12RRL}, suggesting these velocity gradients are consistent with infalling ionised gas.

For these regions, we suggest that the bulls-eye features are infalling  and not contracting because, according to \citet{depree14}, a contracting or collapsing \hii{} region should be distinguished by a decrease in flux over time, not as an inwards velocity gradient or change in size of the \hii{} region. These regions should be monitored over the coming decades to properly distinguish between these cases.

From the moment one maps of these two regions (G302.02 and G336.98 in Figures \ref{fig:G302.02} and \ref{fig:G336.98} respectively), the velocity shifts from the edges to the centres of these \hii{} regions range from $\sim$ 7 to 10 km s$^{-1}$, respectively. These are large on the 3000-4000 au scales of these \hii{} regions and correspond to velocity gradients of 300-500 km s$^{-1}$ pc$^{-1}$.

\begin{table}
\caption{Gaussian fits to H29$\alpha$ emission.} 
\begin{tabular}{rr@{$\pm$}lr@{$\pm$}lrr@{$\pm$}l} 
\hline \hline 
Target & \multicolumn{2}{c}{Amp.} & \multicolumn{2}{c}{Central Velo.} & V$_{\rm LSR}$  & \multicolumn{2}{c}{Line Width} \\ 
  & \multicolumn{2}{c}{(mJy)} & \multicolumn{2}{c}{(km s$^{-1}$)} & (km s$^{-1}$) & \multicolumn{2}{c}{(km s$^{-1}$)}  \\ 
 \hline \hline
G302.02&9.7&0.1&-46.0&0.2&-37.1&11.4&0.2\\ 
G330.28&1.2&0.2&-99.1&1.8&-93.5&12.3&1.8\\ 
G336.98&3.2&0.2&-95.1&0.0&-75.1&13.6&0.0\\ 
G337.63&3.6&0.2&-53.5&0.5&-56.5&9.3&0.5\\ 
G337.84&6.6&1.0&-42.2&1.4&-40.4&8.2&1.4\\ 
G339.11&5.1&0.4&-90.4&1.7&-78.2&13.6&0.0\\ 
\hline \hline 
\label{tab:RRL_gaussians} 
\end{tabular} 
\end{table}

\begin{table}
\caption{Derived HII region properties based on H29$\alpha$ emission. Temperatures derived assuming line widths are thermal, and voigt profile widths are taken to be the greater of half the channel width (0.25 km s$^{-1}$) or the uncertainty on the Gaussian fit in Table \ref{tab:RRL_gaussians}.} 
\begin{tabular}{lr.rrr} 
\hline \hline 
Target & T & \multicolumn{1}{r}{$\Delta v_p$} & $n_e$  &\multicolumn{1}{r}{$M_{\rm{HII}}$}&\multicolumn{1}{r}{S$_c$}\\ & (K) & \multicolumn{1}{r}{(km s$^{-1}$)} & (10$^7$ cm$^{-3}$) &\multicolumn{1}{r}{M$_\odot$}&(mJy)\\\hline \hline 
G302.02 &2800 & 0.25 & $<$1.5 & $<$38.4 & 11\\ 
G330.28 &3300 & 1.8 & $<$11 & $<$482.9 & 2.3\\ 
G336.98 &4100 & 0.25 & $<$1.5 & $<$274.0 & 8.7\\ 
G337.63 &1900 & 0.54 & $<$3.0 & $<$80.2 & 2.7\\ 
G337.84 &1500 & 1.4 & $<$7.7 & $<$220.1 & 3.4\\ 
G339.11 &4100 & 0.25 & $<$1.5 & $<$1206.7 & 1.9\\ 
\hline \hline 
\label{tab:HII_prop} 
\end{tabular} 
\end{table}

The most surprising result of this study is the high percentage (up to 30\%, if the gradient in G339.11 is to be believed) of \hii{} regions detected in \RRL{29} that show infall signatures, when it was expected \citep[from statistical lifetime arguments:][]{Mottram11, Davies11} that the stars powering these \hii{} regions would have rather short (few $\times10^4$ yr)  ionised accretion lifetimes. However, though the Kelvin Helmholtz contraction timescale for single non-accreting sources of these luminosities (log(L) of 4.2 and 4.4, corresponding to $\sim$14 and 17 M$_\odot$) is of the order 10$^{5}$ yrs, if their mass was assembled more slowly with more modest accretion rates ($\sim$10$^{-4}$ M$_\odot$ yr$^{-1}$) then the models predict that the swelling is more modest and such (proto)stars would reach a phase of main-sequence phase already at these sorts of masses \citep[see e.g.][]{Hosokawa10}. The chemical diversity of these two confirmed (and one tentative) infalling regions is also curious. Two are the most chemically rich in our study (G339.11, and G336.98, which, from its radio SED is expected to be one of our youngest \hii{} regions), while the other (G302.02) shows no molecular emission across the \hii{} region (neither in the \RRL{29} nor CS passbands), suggesting that the complex molecules have been destroyed by the ionising radiation.

\newcommand{\IndivCaption}[1]{Continuum, \RRL{29}, and CS emission from {#1}. The top left panel shows the 256 GHz continuum in colour scale in which the 5 GHz emission is shown in red contours; the red contours in the other images represent this same 5 GHz emission for comparison. The yellow contours show the outlines of the \texttt{FellWalker} identified cores.  The bottom left panel shows the \RRL{29} emission (black) with its Gaussian fit. The grey line shows the residuals from the Gaussian fit. In these spectra, the emission has been recentred to the expected rest velocity of the source, thus the velocity shift of the line peak (from zero) represents a difference between the ionised and molecular gas velocities. The top two panels on the right show the moment 0 and moment 1 maps of the \RRL{29} emission (top and bottom), while the bottom two panels on the right show the moment 0 and moment 1 maps of the CS emission (top and bottom). }

\begin{figure*}
\includegraphics[width=0.95\textwidth]{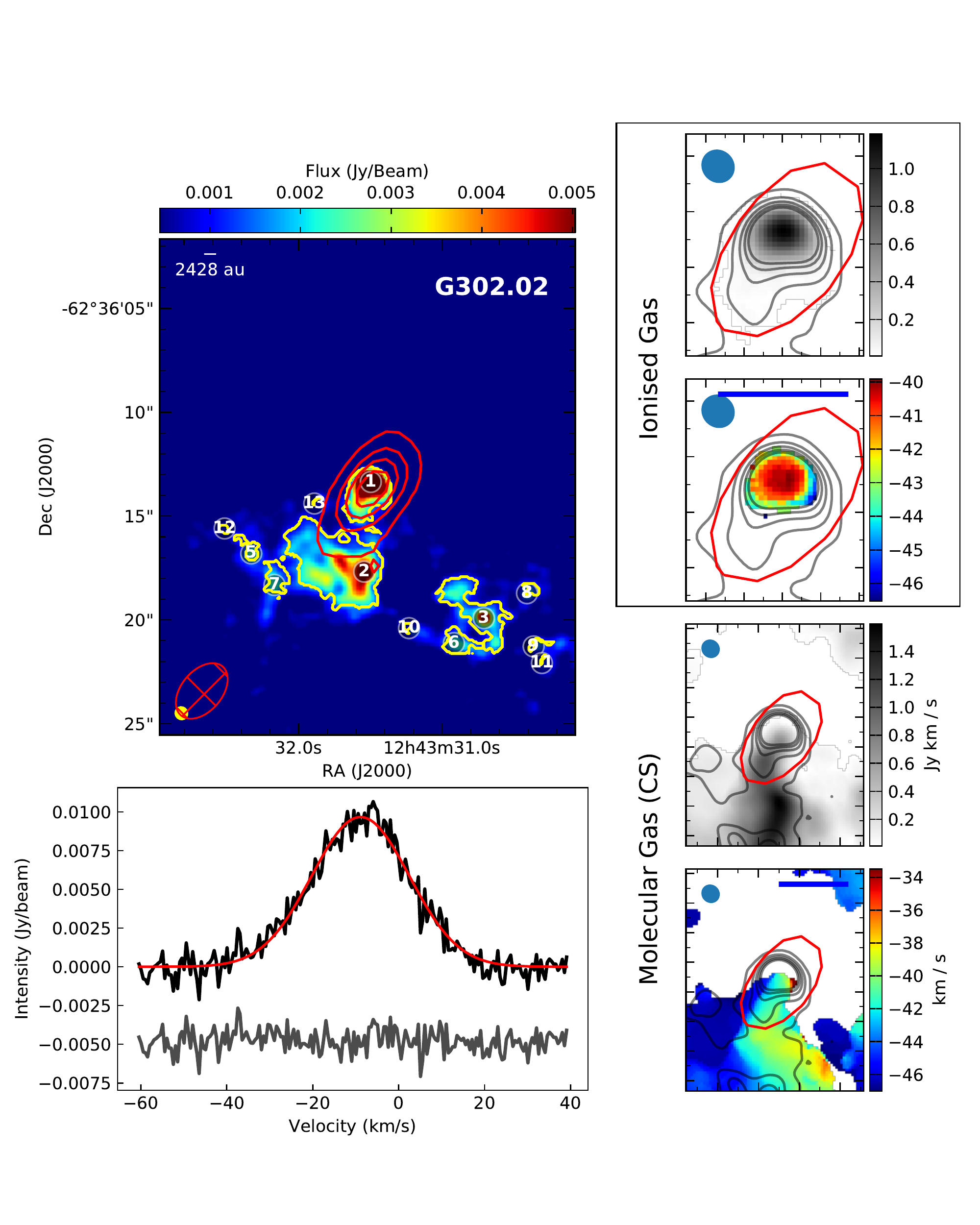}
\caption{\IndivCaption{G302.02}}
\label{fig:G302.02}
\end{figure*}

\begin{figure*}
\includegraphics[width=0.95\textwidth]{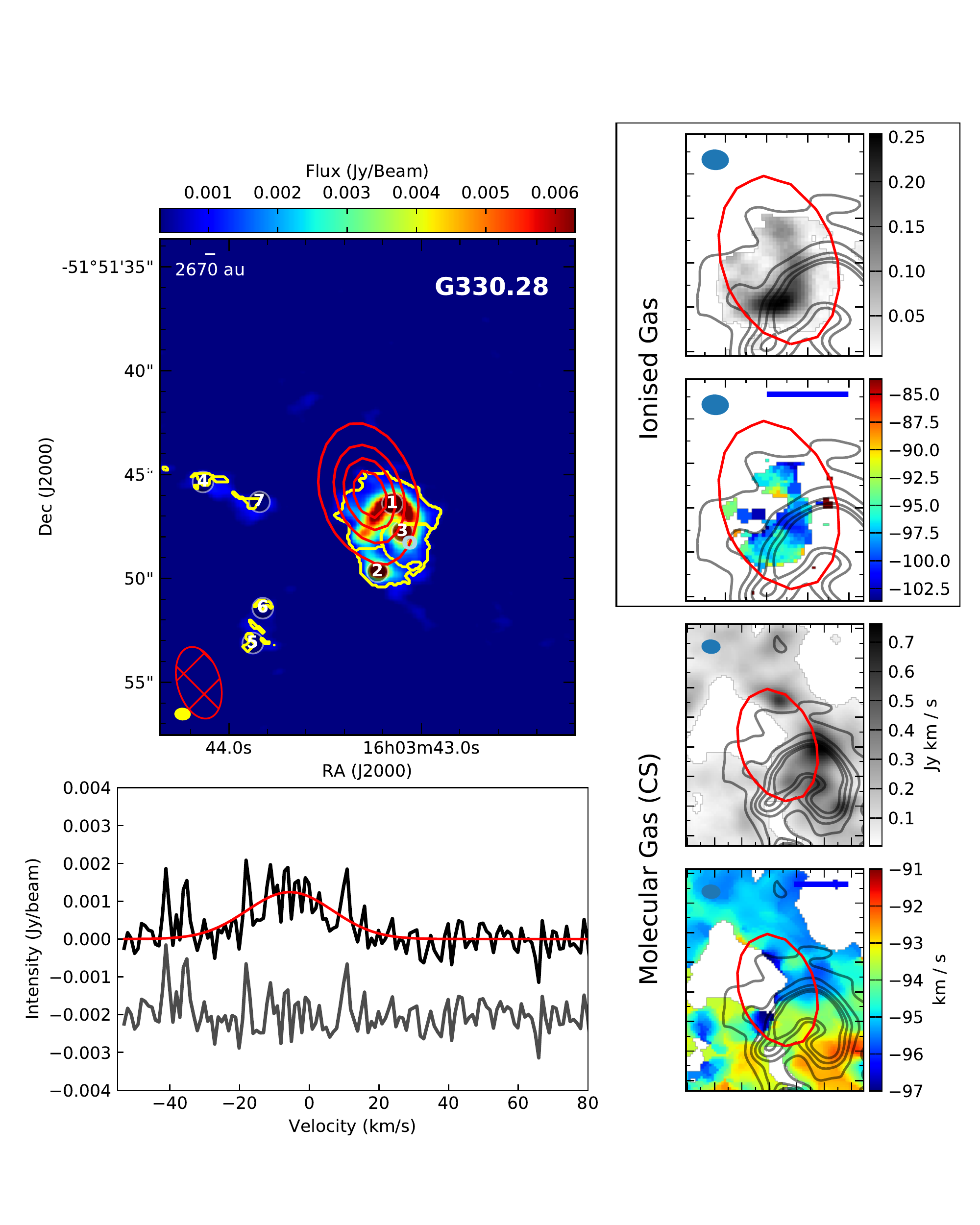}
\caption{\IndivCaption{G330.28}. }
\label{fig:G330.28}
\end{figure*}

\begin{figure*}
\includegraphics[width=0.95\textwidth]{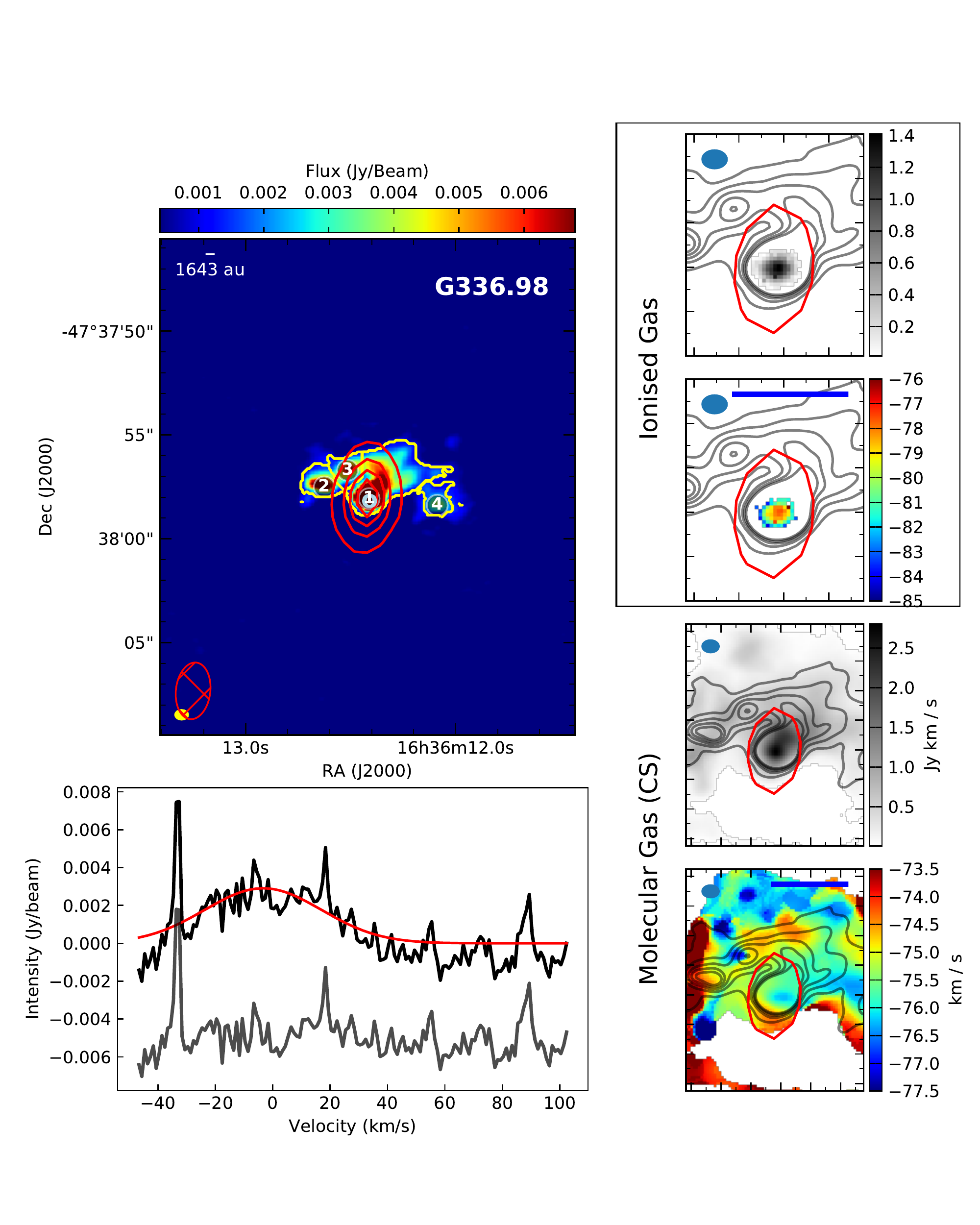}
\caption{\IndivCaption{G336.98} }
\label{fig:G336.98}
\end{figure*}

\begin{figure*}
\includegraphics[width=0.95\textwidth]{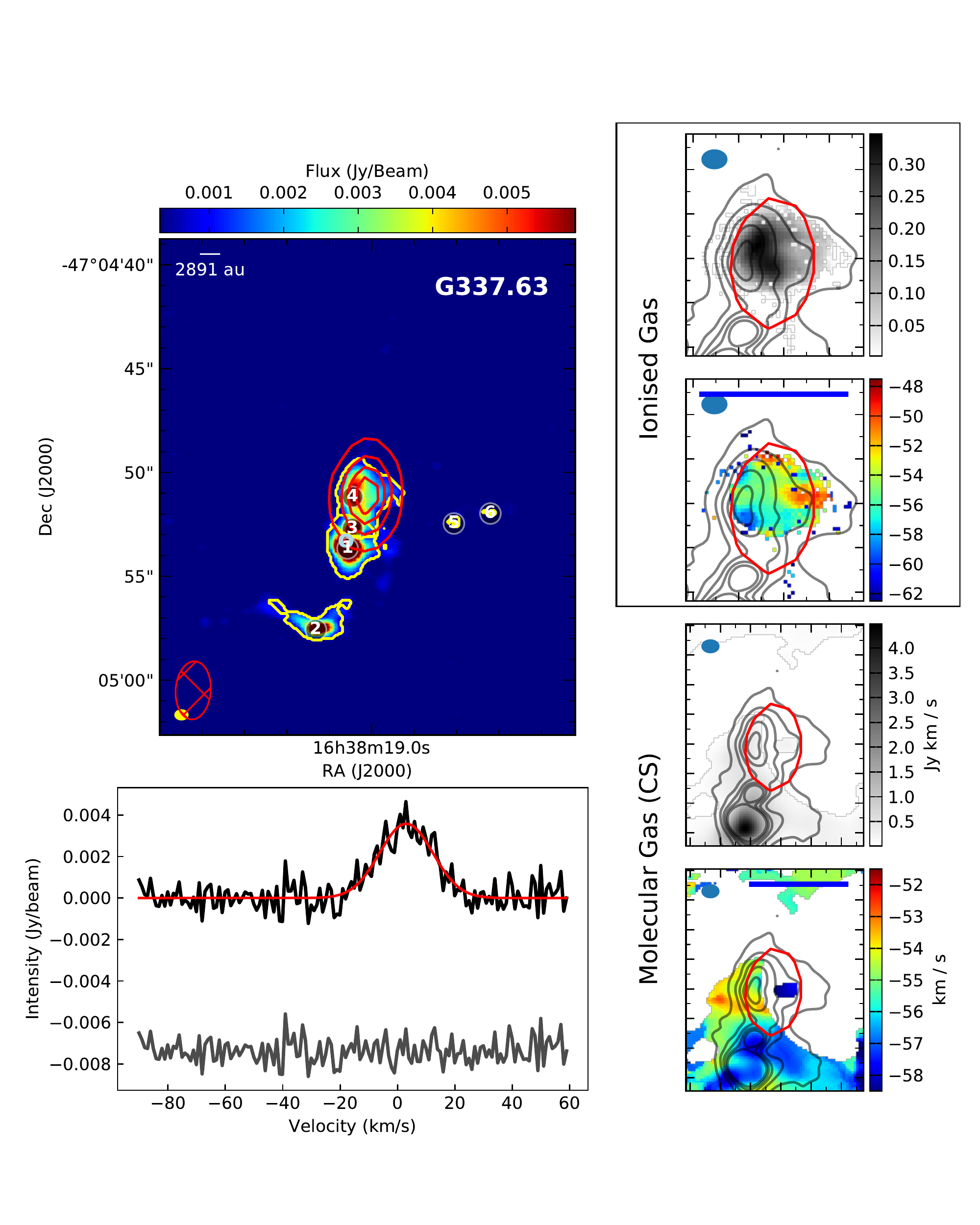}
\caption{\IndivCaption{G337.63}}
\label{fig:G337.63}
\end{figure*}

\begin{figure*}
\includegraphics[width=0.95\textwidth]{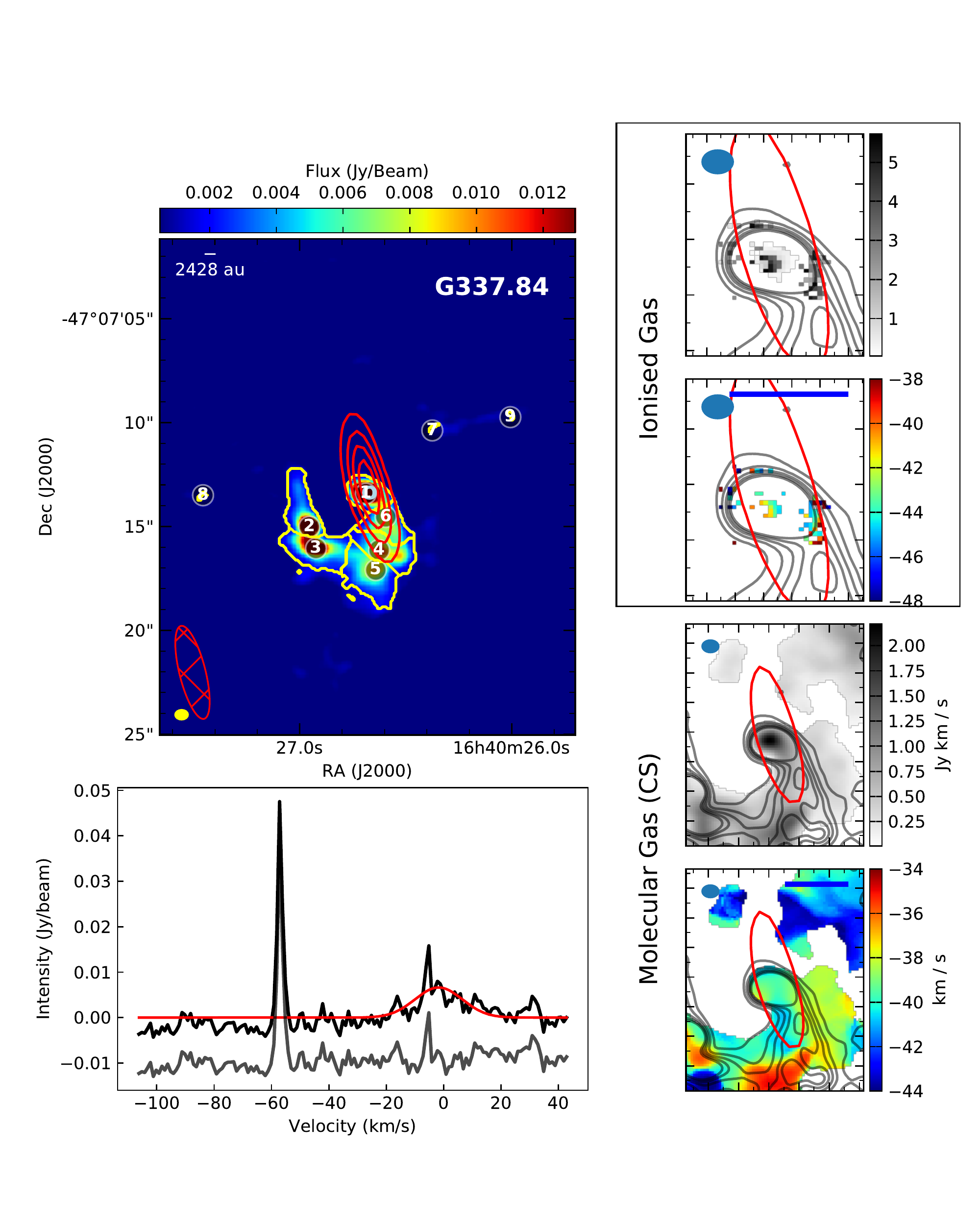}
\caption{\IndivCaption{G337.84}}
\label{fig:G337.84}
\end{figure*}

\begin{figure*}
\includegraphics[width=0.95\textwidth]{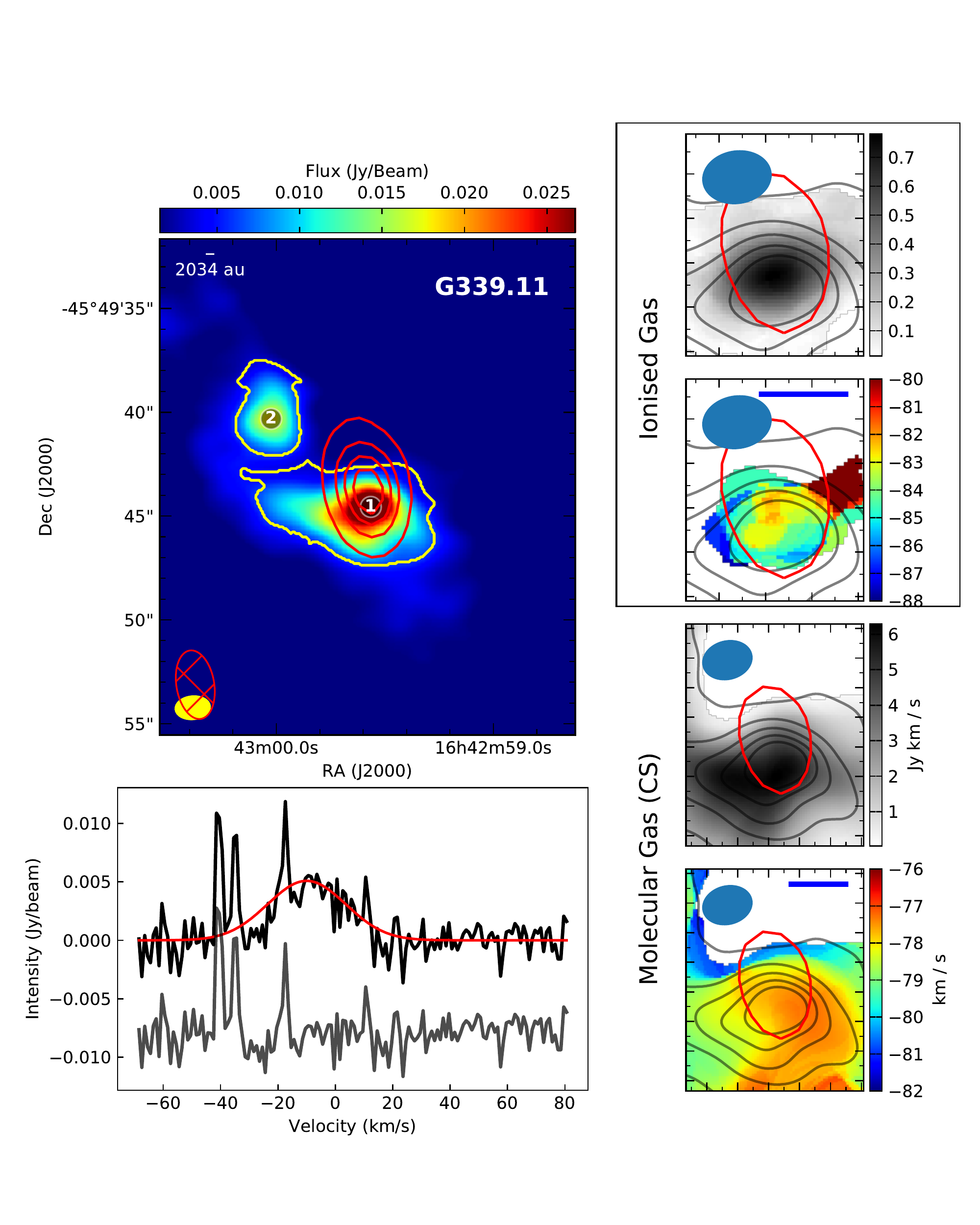}
\caption{\IndivCaption{G339.11}}
\label{fig:G339.11}
\end{figure*}

\newcommand{\IndivCSCaption}[1]{Continuum and CS emission from {#1}. The left panel shows the 256 GHz continuum in colour scale with 5 GHz emission in red contours. The yellow contours show the outlines of the \texttt{FellWalker} identified cores. The two right panels show the moment 0 and moment 1 (top and bottom, respectively) maps of the CS emission focussed on the brightest clump, where the \hii{} region is expected to be (although no \RRL{29} was detected, the 5 GHz continuum shows its location). }

\begin{figure*}
\begin{center}
\includegraphics[width=0.8\textwidth]{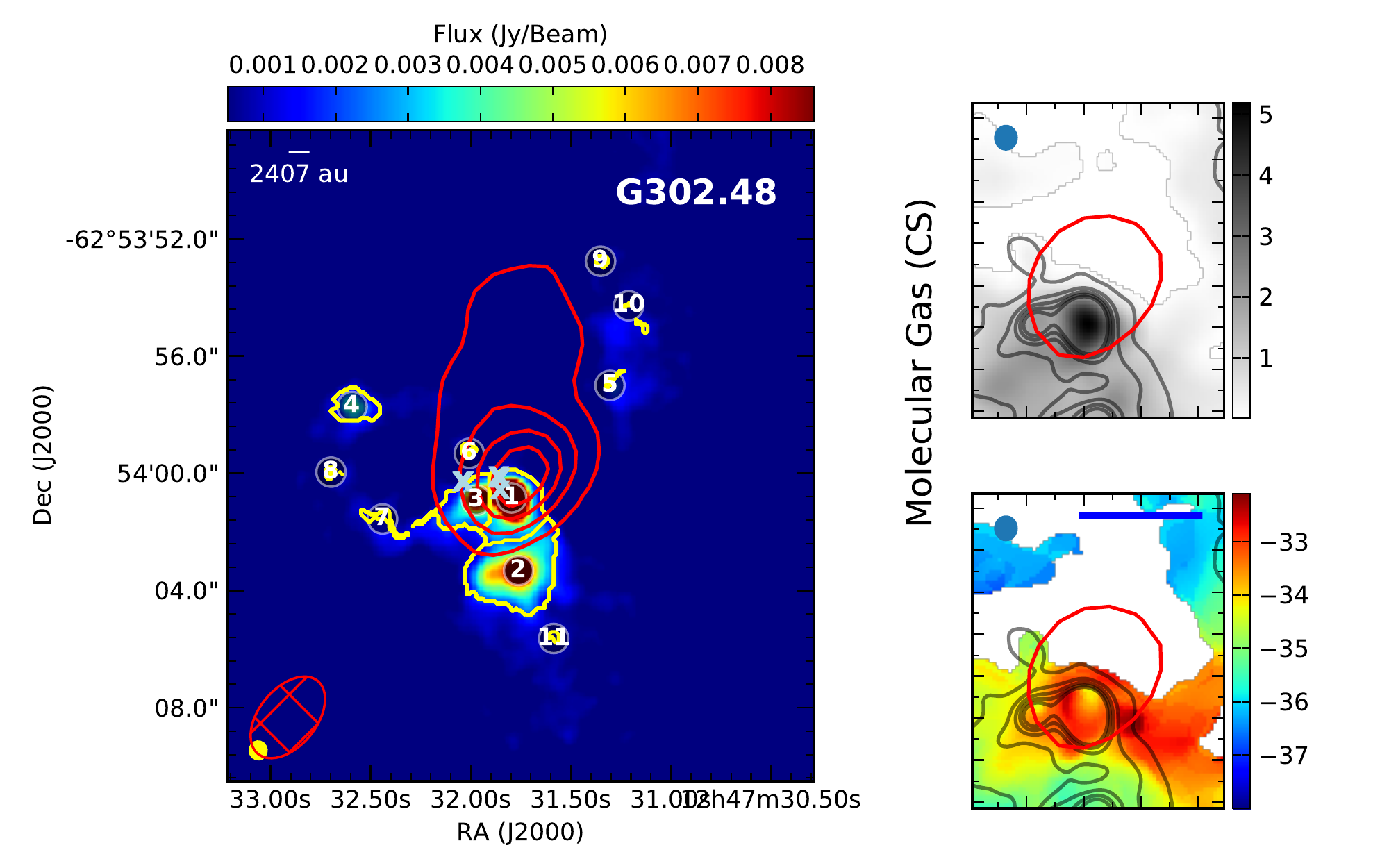}
\end{center}
\caption{\IndivCSCaption{G302.48}}
\label{fig:G302.48}
\end{figure*}

\begin{figure*}
\begin{center}
\includegraphics[width=0.8\textwidth]{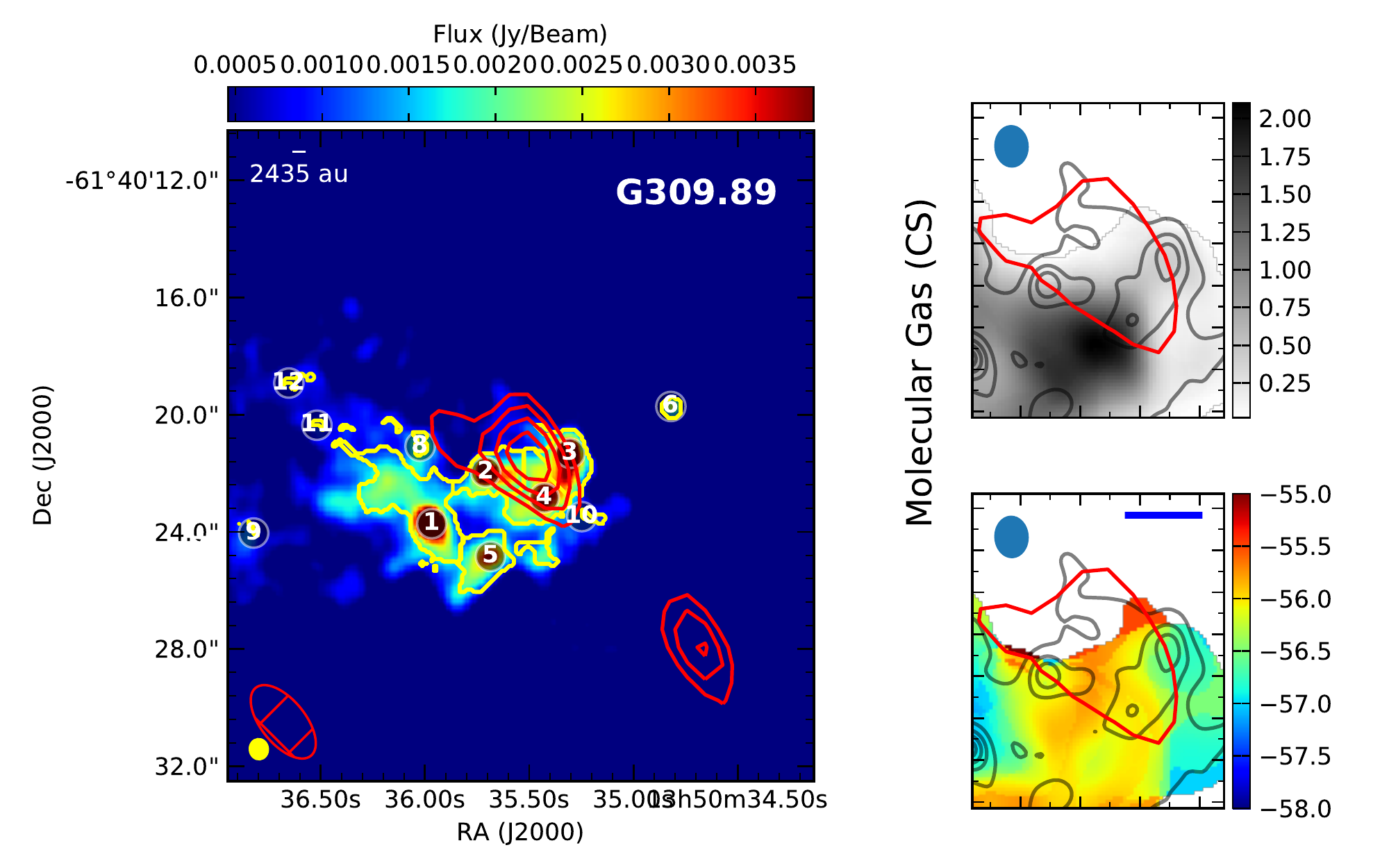}
\end{center}
\caption{\IndivCSCaption{G309.89}}
\label{fig:G309.89}
\end{figure*}

\begin{figure*}
\begin{center}
\includegraphics[width=0.8\textwidth]{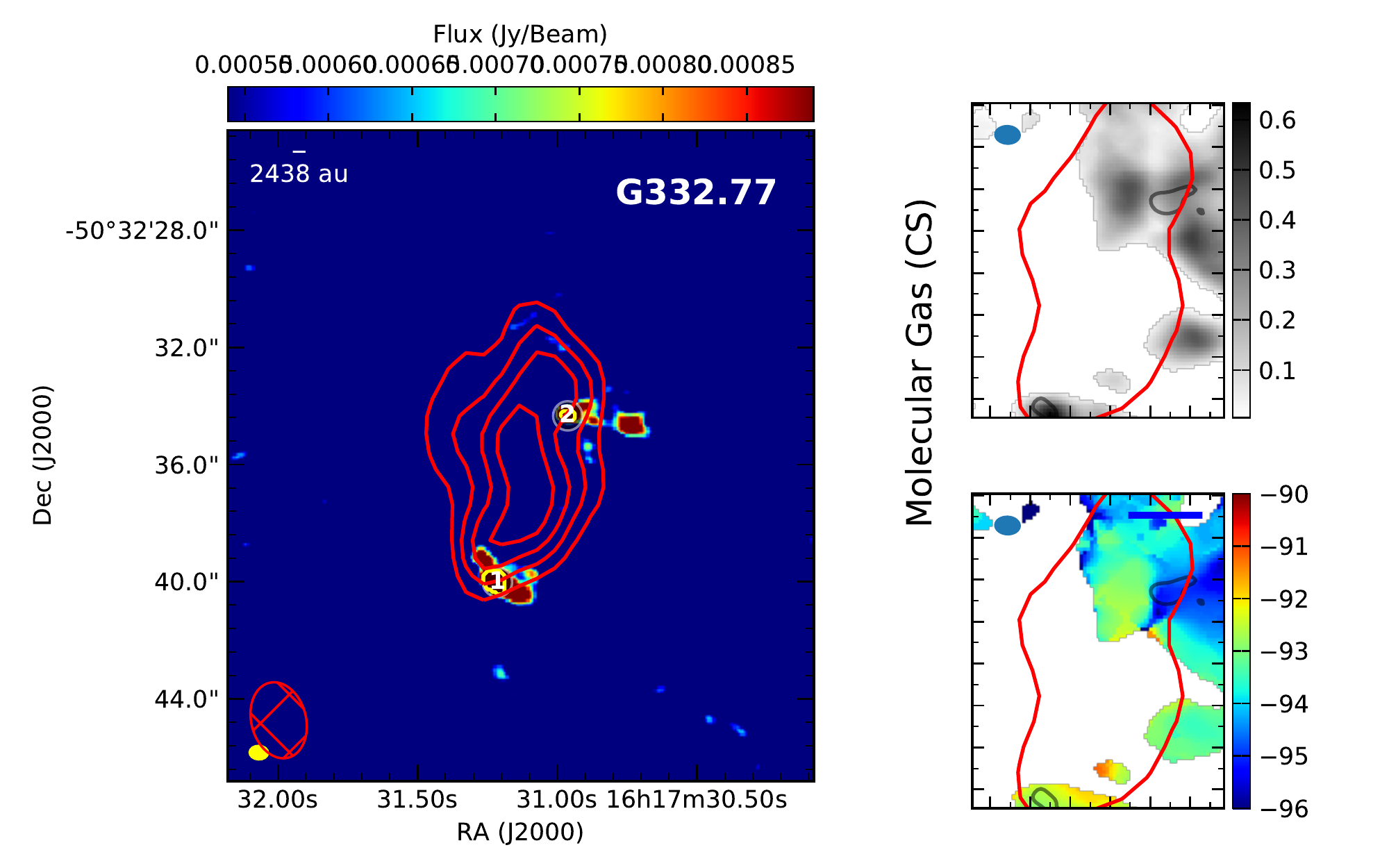}
\end{center}
\caption{\IndivCSCaption{G332.77}}
\label{fig:G332.77}
\end{figure*}

\subsection{Models of ionised gas}
\label{sec:models}

\begin{figure}
\includegraphics[width=\columnwidth]{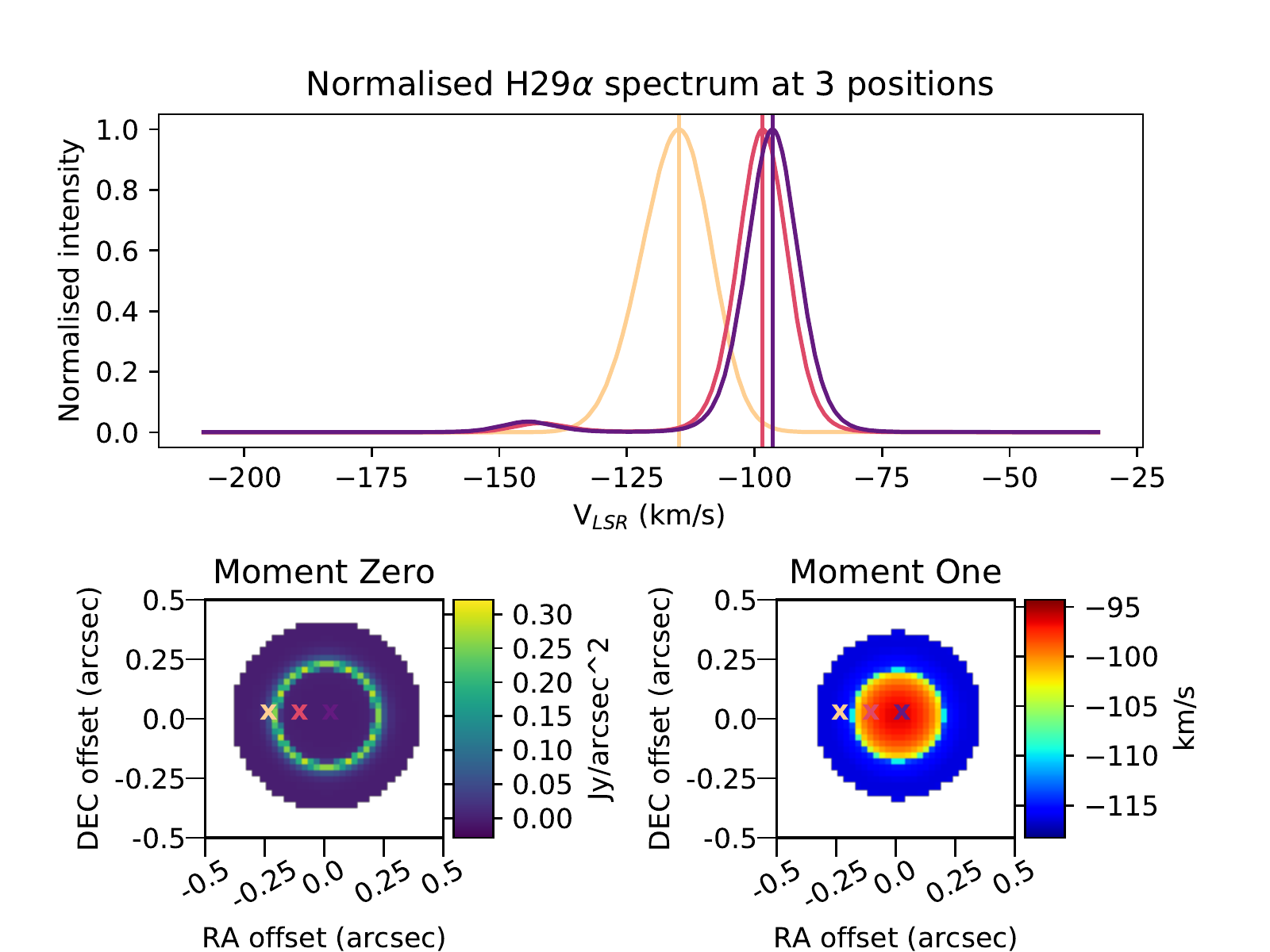}
\caption{Modelled \RRL{29} emission coming from an infalling \hii{} region.  Model taken from \citet{Peters12RRL} and rerun for the \RRL{29} transition.}
\label{fig:infall_model}
\end{figure}

\citet{Peters12RRL} implemented the modelling of RRL line profiles detectable at JVLA and ALMA wavelengths into  \texttt{RADMC-3D} using non-LTE methods to understand line shapes. The test cases they used to demonstrate the code focussed on infalling and outflowing RRL emission, and under which conditions these signatures would be detectable.

These conditions require the observations of the optically thick RRLs to be  of high resolution and sensitivity, and for the intrinsic line widths to be narrow enough for the asymmetries to be spectrally resolved (e.g. pressure broadening does not dominate the spectrum). In Figure 14 of \citeauthor{Peters12RRL}, these authors outlined the parameter space in which line profile variations as a function of position can be attributed to dynamics, and find a parameter space in the top right of their figure where RRL emission is optically thick and the medium thermally supported, thereby requiring high frequency observations of the densest \hii{} regions.

Our observations ($\nu\sim256$ GHz, $n_e< 5\times10^7$ cm$^{-3}$) fall into this parameter space, and indeed, we do find   these line asymmetries in some of our regions.

We reran their models for infalling gas as traced by \RRL{29}. These model results are shown in Figure \ref{fig:infall_model}, where we present the zeroth and first moment maps along with normalised spectra taken at radial points along the optically thick to thin transition (outwards).  The first moment map clearly shows red-shifted emission in the optically thick region.  Because the emission is optically thick, we are probing the near side of the \hii{} region and the fact that the emission is red-shifted indicates the gas is moving away from us; it is falling inwards onto the \hii{} region.  A qualitative comparisons to our data then suggests that the bulls-eye patterns we see in G302.02 and G336.98 are indicative of infall motions onto/through the \hii{} region.

\subsection{Molecular gas dynamics}
\label{subsec:molecular}

\subsubsection{\texttt{XCLASS} fitting: Molecular results}
\label{sec:CH3CCH_results}

There were two molecular species fit during the \texttt{XCLASS} modelling (See Appendix \ref{sec:remove_molecules}) that were parameterised well enough to constrain some of the properties of the warm molecular gas surrounding the \hii{} regions. Towards G336.98, both \methacet{} (Figure \ref{fig:CH3CCH_G336.98}) and CH$_3$OCHO were well constrained with the best fit models giving temperatures of $\sim 70-80$ K and column densities of 3-10$\times10^{14}$ cm$^{-2}$. The \methacet{} shows a velocity gradient across the area surrounding the \hii{} region (roughly orientated E-W), which is different from the infall signature seen in the \RRL{29}.  The CH$_3$OCHO does not show a velocity gradient. There are two reasons why this could be the case; either there is no velocity gradient or, more likely the fit was not good enough to identify a velocity gradient since the $\chi^2$ values for CH$_3$OCHO were much larger than for \methacet{}.  With \methacet{}, there were five transitions available to constrain the velocity gradient and with CH$_3$OCHO, there is only one. Thus, it becomes easier for velocity, density, and temperature gradients to be mis-identified.

\methacet{} was also well constrained in G302.48 (T$\sim70$ K, N $\sim 10^{15}$ cm$^{-2}$), but we can draw no comparison to the ionised gas dynamics as \RRL{29} was not detected in this region.

CH$_3$OCHO was detected in G337.84 (where a very faint detection of \RRL{29} was also made) with suggestions of higher temperatures and column densities (T$\sim300$ K, N$\sim10^{15}$ cm$^{-2}$).  Its velocity structure can be described as saddle-like and is reminiscent of the CS velocity patterns presented for this target below in Section \ref{sec:CS}, and in Figure \ref{fig:G337.84}.

\subsubsection{Comparison to CS emission}
\label{sec:CS}

There were a number of molecular species observed along with \RRL{29} in our observing set-up and the analysis of the larger scale molecular dynamics in these regions will be discussed in a forthcoming paper (Klaassen et al. in prep).  However, it is instructive to put the RRL emission into context with its surrounding molecular gas. In this light, we present a brief summary of the CS (J=5-4) dynamics surrounding each \hii{} region.

Integrated intensity and intensity weighted velocity (zeroth and first moment) maps are presented in Figures \ref{fig:G302.02} - \ref{fig:G332.77} (right panels labelled as `molecular'). As can be seen from these maps, the velocity ranges over which the CS emits are much smaller than those of the \RRL{29}.

For the \hii{} regions with cometary morphologies, unsurprisingly the molecular emission also appears to have a somewhat cometary structure.  In the two cases presented here, we note that the CS emission generally tends to be detected around the \hii{} region, rather than coincident with it. This dearth of molecular emission suggests  either that the ionised gas has cleared away a region for it to expand (i.e. a champagne flow) through the lower density medium, or that the \hii{} region is moving through the molecular gas and creating a bow of molecular emission at its head.  From the current data, we cannot break this degeneracy.

In the cases of the regions with suggested infalling motions in the ionised gas, the CS emission does not show much correlation between regions. Around G302.02 the CS shows evidence for some kind of velocity gradient roughly east to west below the \hii{} region. For the other region with a suggestion of ionised infall (G336.98), in the \RRL{29}, there is no equivalent collimated structure in the CS. In both cases, however, the \hii{} region does, again, appear to be near the edges of the molecular emission.

\section{Conclusions}
\label{sec:conclusions}

We have conducted a study of the mm emission from nine high-mass star-forming regions associated with young \hii{} regions.  We show that these regions are clustered (or at least clumpy) and that the separations between cores are more consistent with the thermal Jeans lengths than the turbulent Jeans lengths given the pre-cursor densities and temperatures of the cores.  We determined the masses of each of the identified cores, and find that, at most, a few of these cores in each region  have masses great enough to provide enough mass for a high-mass star, if they have low temperatures.

We find evidence for \RRL{29} emission in six out of nine regions, and that, in many of these regions, the \RRL{29} spectrum is contaminated by molecular emission that needs to be disentangled to properly quantify the RRL emission.   None of the \RRL{29} profiles show any indications of pressure broadening, which puts upper limits on the electron densities of these \hii{} regions at roughly 10$^7$ cm$^{-3}$. The \hii{} regions with \RRL{29} detections can broadly be separated into two types: those defined spatially as cometary (G330.28 and G337.63), and those defined by their bulls-eye dynamics (G302.02, G336.98, and potentially G339.11) indicative of infalling motions.

Of our sample of nine, the regions expected to have the youngest \hii{} regions (from their radio spectral indices) show evidence for infall motions in the \RRL{29} emission. This could be indicative of late stage accretion onto the central star(s) despite the presence of \hii{} regions.  In these cases, we find evidence both for the infalling \hii{} region to be shrouded in its chemically rich hot core, and for this infalling region to have emerged from this hot core.  For those \RRL{29} detections that appear cometary (both morphologically and in their velocities), we find a much stronger correlation with the emission of the surrounding molecular gas; there is very little molecular gas at the tail and there is a bow in the molecular gas ahead of the \hii{} region.  The                 question of whether the bow of molecular emission is caused by the \hii{} region moving through the medium or is the reason the cometary morphology exists (e.g. it is a champagne flow through a low density cavity) requires further study and comparison with models to disentangle.

\bibliographystyle{aa}

\begin{acknowledgements}
This research made use of NASA's Astrophysics Data System; the SIMBAD database, operated at CDS, Strasbourg, France; APLpy, an open-source plotting package for Python hosted at http://aplpy.github.com; Astropy, a community-developed core Python package for Astronomy \citep{astropy}; pyspeckit, an open-source spectral analysis and plotting package for Python hosted at http://pyspeckit.bitbucket.org. This paper makes use of the following ALMA data: ADS/JAO.ALMA\#2013.1.00327.S. ALMA is a partnership of ESO (representing its member states), NSF (USA), and NINS (Japan), together with NRC (Canada), NSC and ASIAA (Taiwan), and KASI (Republic of Korea), in cooperation with the Republic of Chile. The Joint ALMA Observatory is operated by ESO, AUI/NRAO and NAOJ. JCM and HB acknowledge support from the European Research Council under the European Community's Horizon 2020 framework programme (2014-2020) via the ERC Consolidator grant 'From Cloud to Star Formation (CSF)' (project number 648505). TP acknowledges the {\em Deutsche Forschungsgemeinschaft (DFG)} for funding through the SPP 1573, "The Physics of the Interstellar Medium". RK acknowledges financial support via the Emmy Noether Research Group on Accretion Flows and Feedback in Realistic Models of Massive Star Formation funded by the German Research Foundation (DFG) under grant no. KU 2849/3-1. 
\end{acknowledgements}

\onecolumn
\begin{appendix}

\section{Capturing the flux on large scales}
\label{sec:short_spacing}

\subsection{Combining multi-configuration datasets}

For the continuum datasets, the relevant (i.e. line free) channels were averaged together to speed processing, and then the measurement sets from the individual array configurations were concatenated together. This single concatenated dataset was then Fourier transformed into the image plane using the \verb+clean+ command.  Similarly, for the line datasets, we found that splitting out the relevant channels of the given spectral window (i.e. the channels within 100 km s$^{-1}$ of the rest velocity), concatenating the configurations, and inverting the concatenated dataset gave the best signal-to-noise ratios in the final image plane maps.

As described further below, we quantified whether the ACA data was required for imaging. In terms of \RRL{29} emission, we found no significant differences (< 1$\sigma$)  in the emission  profiles between combining only  12 m configuration data and combining them with the ACA data as well in the target with the brightest \RRL{29} emission (G302.02).  For the continuum emission, we found that more flux was recovered when incorporating the ACA data.  Comparing the integrated fluxes in our 26$''$ primary beam to those from the ATLASGAL 350 GHz emission (averaged over a 35$''$ beam), we found that with the combination of 12 m and ACA observations, we recovered all of the flux in these regions with our observations, taking into account the different frequencies of the observations. The largest angular scale to which our observations were sensitive was comparable to our primary beam, and our achieved synthesised beams and sensitivities are listed in Table \ref{tab:obs_params}.

\begin{table*}
\caption{Calibrators and Observing Parameters for ALMA project 2013.1.00327.S}
\begin{tabular}{llllllll}
\hline \hline
&Time on each& \multicolumn{3}{c}{Calibrators} & & \multicolumn{2}{c}{Baseline Lengths}\\ \cline{3-5} \cline{7-8}
Execution\tablefootmark{a}  & Science Source & Phase & Bandpass & Flux & Observing Date & Max & Min\\
 & (min) &&&&&(m) &(m)\\
 \hline
\\
\multicolumn{8}{c}{Targets near G305  (3 regions)}\\
\\
\hline
\multicolumn{8}{c}{{\bf extended 12m configuration} (pwv = 1.45 mm)}\\
\hline
X230 & 9 & J1254-6111 & J1427-4206 & J1107-448 & 2015-06-04 & 784 & 21.4\\
\hline
\multicolumn{8}{c}{{\bf compact 12m configuration} (pwv = 3.82 mm)}\\
\hline
X2474 & 4 & J1254-6111 & J1107-4449 & Callisto & 2015-01-15 & 349 & 15.1\\
\hline
\multicolumn{8}{c}{{\bf ACA} (pwv = 1.52 mm)}\\\hline
X610 & 8 & J1308-6707 & J1427-4206 & Titan & 2014-06-07 & 48 & 9.1\\
X1ff & 8 & J1308-6707 & J1427-4206 & Titan & 2014-06-07 & 48 & 9.1\\
X610 & 8 & J1308-6707 & J1427-4206 & Titan & 2014-06-07 & 48 & 9.1\\
X1ff & 8 & J1308-6707 & J1427-4206 & Titan & 2014-06-07 & 48 & 9.1\\
\hline
\\
\multicolumn{8}{c}{Targets near G335 (6 regions)}\\
\\
\hline
\multicolumn{8}{c}{{\bf extended 12m configuration} (pwv = 1.36 mm)}\\
\hline
X15f4 & 8 & J1617-5848 & J1617-5848 & Titan & 2015-06-04 & 784 & 21.4\\
\hline
\multicolumn{8}{c}{{\bf compact 12m configuration} (pwv = 1.96 mm)}\\
\hline
X2ade & 4 & J1617-5848 & J1617-5848 & Ceres & 2015-01-03 & 349 & 15.1\\
\hline
\multicolumn{8}{c}{{\bf ACA} (pwv = 0.84 mm)}\\
\hline
X6823 & 5 & J1617-5848 & J1427-4206 & Titan & 2015-06-07 & 49 & 8.9\\
X1f1 & 5 & J1617-5848 & J1617-5848 & J1613-586 & 2014-07-29 & 49 & 8.9\\
X1804\tablefootmark{b} & 1 & J1617-5848 & J1617-5848 & Titan & 2015-04-22 & 48 & 9.1\\
Xd0 & 5 & J1617-5848 & J1617-5848 & Mars & 2014-07-27 & 49 & 8.9\\
Xa6e & 5 & J1617-5848 & J1617-5848 & J1613-586 & 2014-07-28 & 49 & 8.9\\
X1e58 & 5 & J1617-5848 & J1924-2914 & Titan & 2015-06-03 & 49 & 8.9\\
X1ae0 & 5 & J1617-5848 & J1617-5848 & Titan & 2015-06-03 & 91 & 8.9\\
\hline
\end{tabular}
\tablefoot{
\tablefoottext{a}{The execution ID is the end portion of the unique filename given to every ALMA observation.}
\tablefoottext{b}{G336.98 and G339.11  were not included in this execution}
}
\label{tab:calibrators}
\end{table*}

G302.02 appears to be the target with both the largest scale continuum and \RRL{29} emission, and so to show that we have captured all of the flux for all of our targets, we use this region as an example to show how well the data combinations worked.

\subsection{Continuum emission}

\begin{figure*}
\begin{center}
\includegraphics[width=0.9\textwidth]{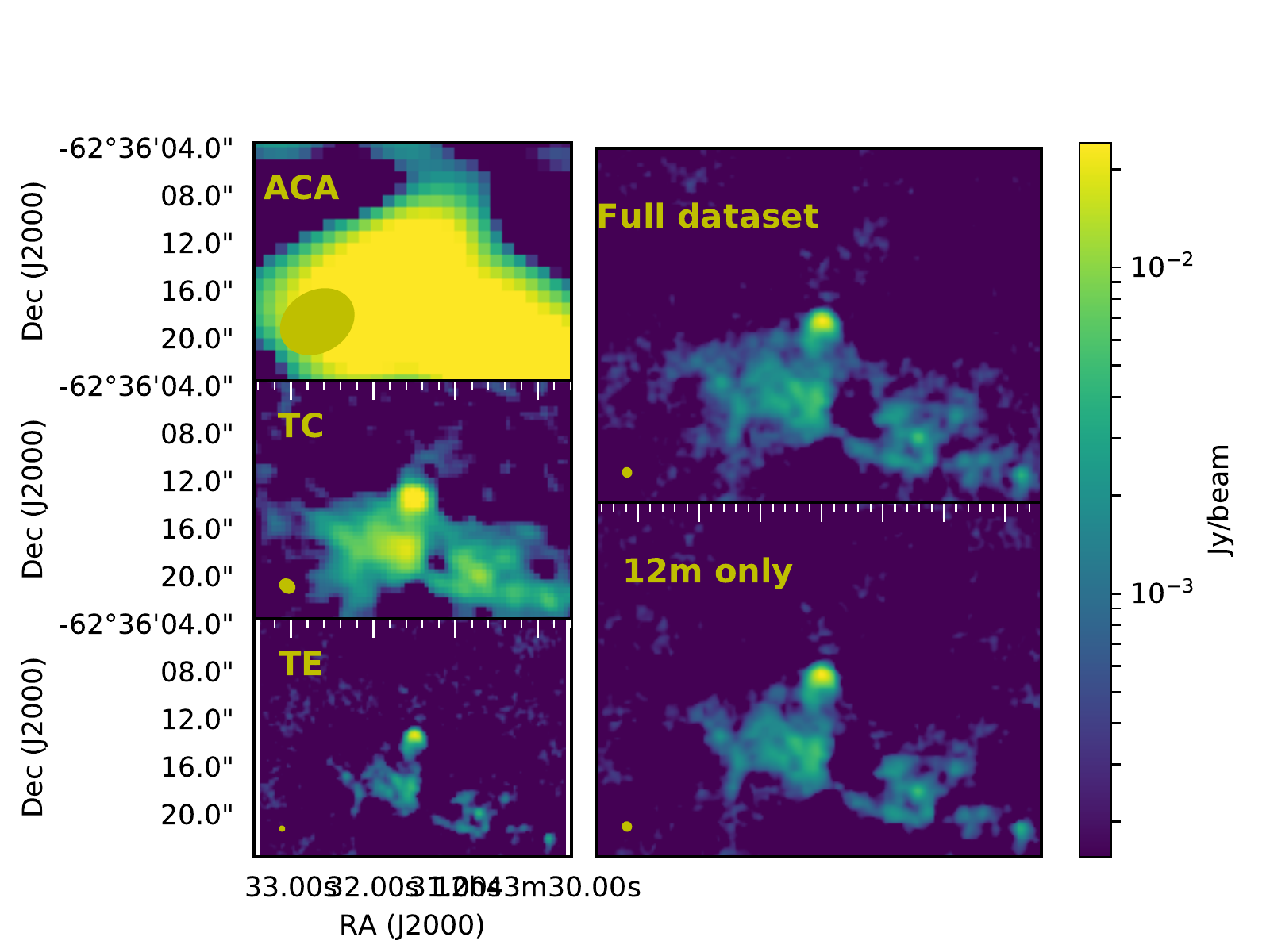}
\end{center}
\caption{{\bf Continuum data combination for G302.02:} The left three panels show the data obtained with each individual antenna configuration; the two 12 m configurations are labelled as TC and TE for the compact and extended configurations, respectively. The right panels show the combined datasets. The bottom right panel shows 12 m data only, while the top right panel shows the full continuum dataset. The integrated fluxes of the two right panels vary by 20\%, although their peak fluxes are relatively consistent with each other.  The scaling is the same for all five plots, as shown at the right.}
\label{fig:cont_ACA}
\end{figure*}

The ACA only images of the continuum emission showed that the continuum was resolved, suggesting that the 12 m data alone would be insufficient to capture the full continuum flux.  The left panels of Figure \ref{fig:cont_ACA} show the continuum emission captured by the two 12 m configurations and that of the ACA for G302.02, which is the target with the most extended emission.  The right panels of  Figure \ref{fig:cont_ACA} shows the combination of 12 m only data (bottom), and combination of 12 m and ACA data (top) for comparison. The colour scale is the same for all panels, shown with logarithmic scaling to highlight low-lying emission, and is tied to the peak flux of the fully combined (12 m+ACA) flux scale.  The peak flux densities in the two right panels (of the combined data) are very similar (to within 3 sigma of each other at 24.1 and 23.9 mJy/beam for the full and 12 m only data, respectively), however in the 12 m only dataset, we recover only 80\% of the flux present in the full (i.e. including ACA) dataset.

\begin{figure*}
\begin{center}
\includegraphics[width=0.8\textwidth]{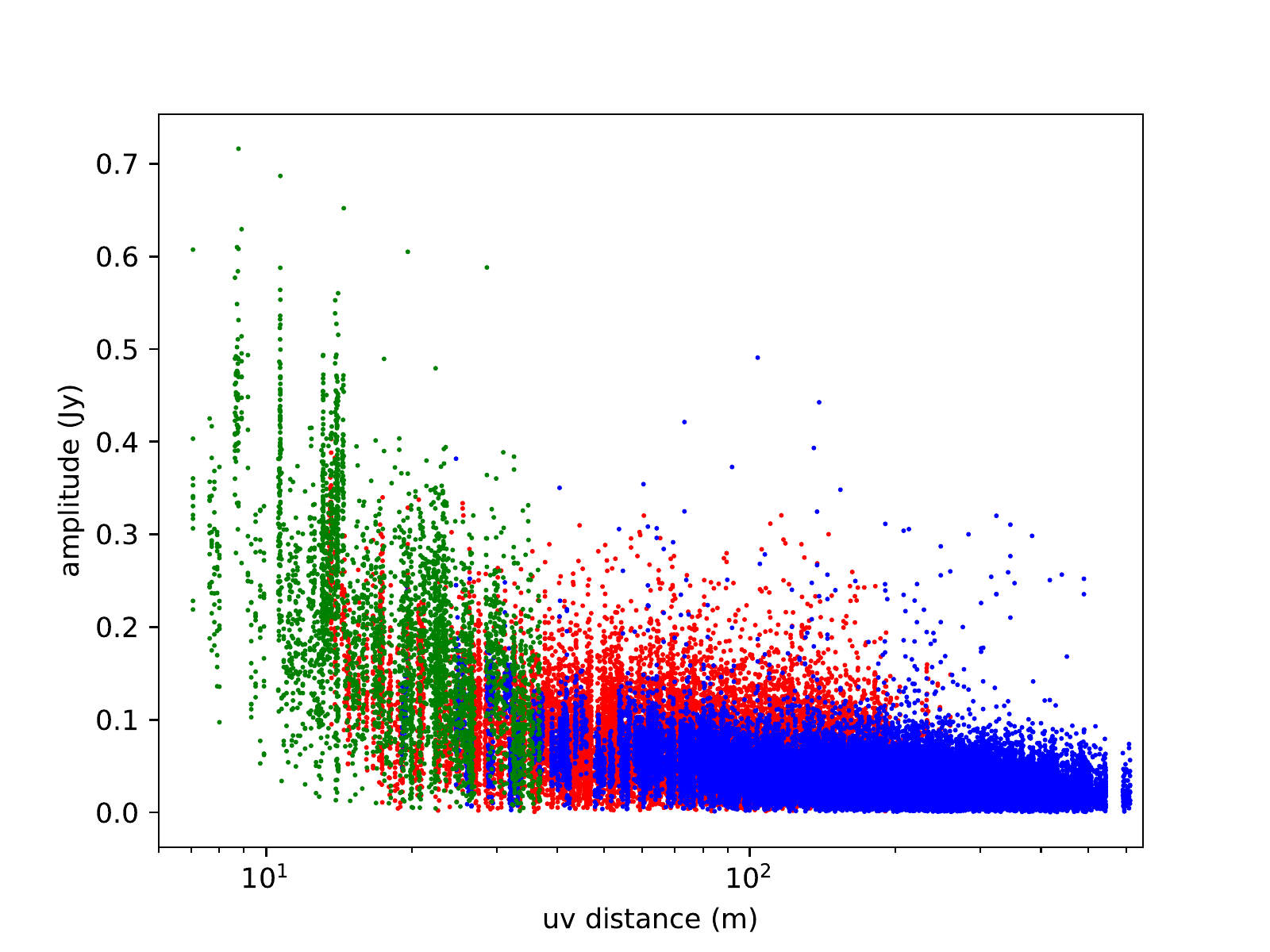}
\end{center}
\caption{Continuum emission in the three configurations: ACA (green), compact 12m (red), and extended 12m (blue). The amplitudes are generally consistent with each other, given the rms noise limits in each map, and show the same trends with uv distance (e.g. higher amplitudes at lower spatial frequencies - larger size scales).}
\label{fig:uvdist}
\end{figure*}

Figure \ref{fig:uvdist} shows how the amplitude of the emission varies as a function of uv distance. Of note here is the overlapping uv distances of the three configurations and generally consistent fluxes between configurations.  The data are sampled well enough in the overlap region to allow for proper combination of the datasets.

\subsection{\RRL{29} emission}

\begin{figure*}
\includegraphics[width=0.8\textwidth]{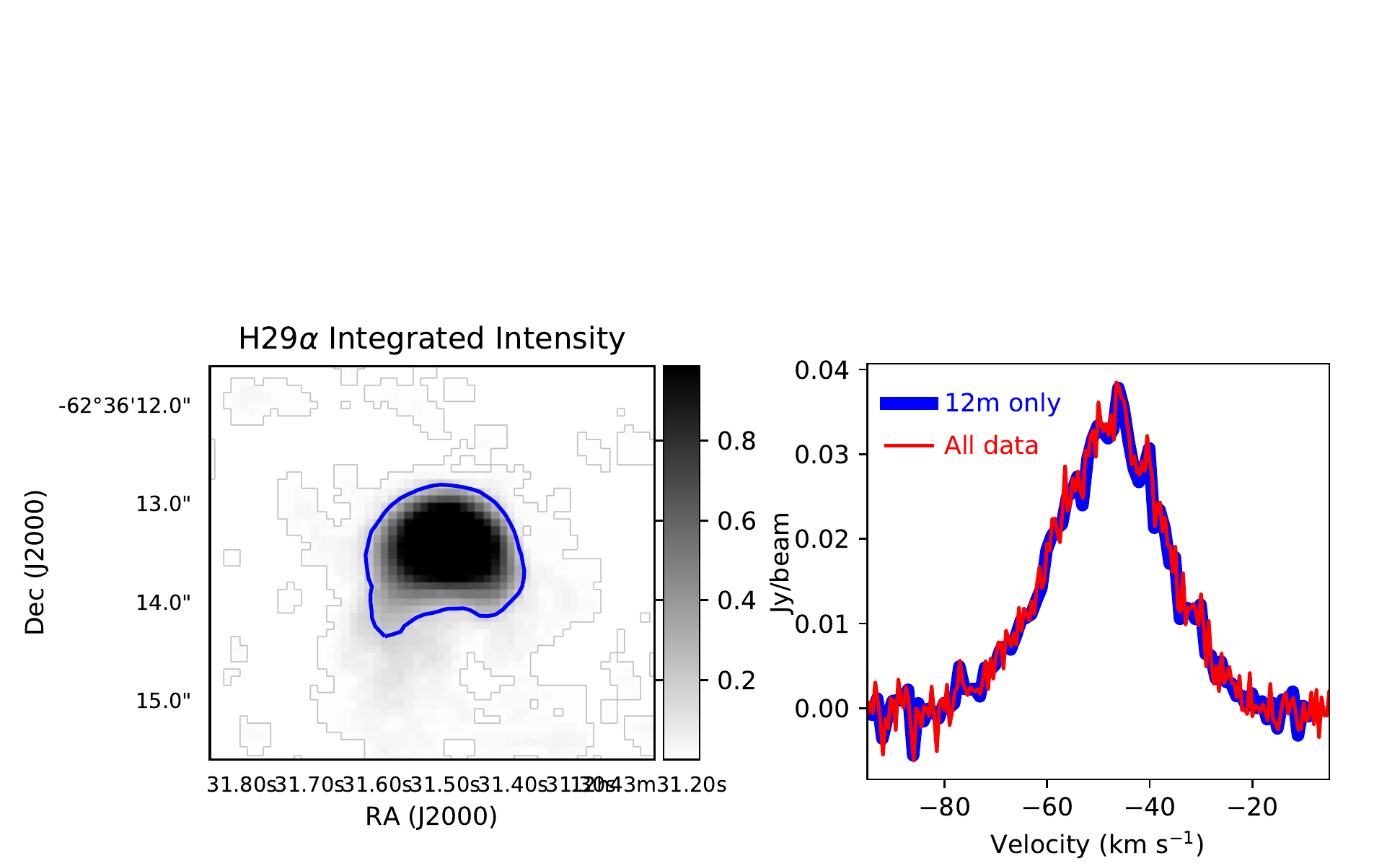}
\caption{{\bf \RRL{29} Data combination:} Comparison between 12 m only and full dataset for capturing the \RRL{29} flux. {\bf Left:} Integrated intensity of \RRL{29} towards G302.02, where the blue contour shows the 10\% power contour of the emission. The map was created with a 3$\sigma$ clipping mask and the greyscale is limited to 50\% power to highlight the low-lying emission. {\bf Right:} \RRL{29} spectra integrated within the 10\% contour shown on the left. The blue (thick) line shows the 12 m only data, while the red (thin) line shows the combination of 12 m and ACA data. The lack of significant differences between the spectra suggests that the ACA is not required to capture all of the flux from the RRLs.}
\label{fig:H29a_comparison_spectra}
\end{figure*}

\RRL{29} was detected in most of our sources and on small enough scales to not require the use of ACA data. It was only detected in the stand-alone ACA data for G302.02, which is the target with the brightest and most extended H29a emission. However, as shown in Figure \ref{fig:H29a_comparison_spectra}, the 12 m array data recover all the flux so it was not necessary to combine this data with the ACA data.

\begin{longtab}
\begin{longtable}{ccccccccccc}
\caption{Clump positions, Areas, Derived Fluxes, Masses, and Column Densities }\\
\hline \hline
ID Number & Peak RA & Peak DEC & Area & Flux & \multicolumn{3}{c}{Clump Mass (M$_\odot$)} & \multicolumn{3}{c}{Log(Column Density) } \\
 & (h:m:s) & (d:m:s) & (arcsec$^{2}$) & (mJy)  & (20 K) & (50 K) & (100 K) & (20 K) & (50 K) & (100 K)\\
\hline \hline
\endfirsthead
\caption{continued.}\\
\hline\hline
ID Number & Peak RA & Peak DEC & Area & Flux & \multicolumn{3}{c}{Clump Mass (M$_\odot$)} & \multicolumn{3}{c}{Log(Column Density) } \\
 & (h:m:s) & (d:m:s) & (arcsec$^{2}$) & (mJy)  & (20 K) & (50 K) & (100 K) & (20 K) & (50 K) & (100 K)\\
\hline
\endhead

\hline \multicolumn{7}{c}{G302.02 (1.00e+07 cm$^{-3}$, 70.0 K)}\\\hline
1 &12:43:31.50 &-62:36:13.36 &4.63 &62.08 &25.37 &8.37 &3.94 &24.41 &23.93 &23.6  \\
2 &12:43:31.54 &-62:36:17.68 &13.45 &81.48 &33.29 &10.99 &5.17 &23.79 &23.31 &22.98  \\
3 &12:43:30.71 &-62:36:19.92 &5.22 &23.92 &9.77 &3.23 &1.52 &23.77 &23.29 &22.96  \\
4 &12:43:29.87 &-62:36:22.00 &0.94 &4.57 &1.87 &0.62 &0.29 &23.62 &23.14 &22.81  \\
5 &12:43:32.33 &-62:36:16.80 &0.75 &3.01 &1.23 &0.41 &0.19 &23.45 &22.97 &22.64  \\
6 &12:43:30.92 &-62:36:21.12 &1.5 &6.01 &2.46 &0.81 &0.38 &23.43 &22.95 &22.62  \\
7 &12:43:32.17 &-62:36:18.32 &1.39 &5.57 &2.28 &0.75 &0.35 &23.41 &22.92 &22.6  \\
8 &12:43:30.41 &-62:36:18.72 &0.38 &1.29 &0.53 &0.17 &0.08 &23.32 &22.84 &22.51  \\
9 &12:43:30.36 &-62:36:21.28 &0.3 &0.83 &0.34 &0.11 &0.05 &23.18 &22.7 &22.37  \\
10 &12:43:31.23 &-62:36:20.40 &0.16 &0.43 &0.17 &0.06 &0.03 &23.12 &22.64 &22.31  \\
11 &12:43:30.30 &-62:36:22.08 &0.09 &0.23 &0.09 &0.03 &0.01 &23.09 &22.61 &22.28  \\
12 &12:43:32.52 &-62:36:15.60 &0.2 &0.52 &0.21 &0.07 &0.03 &23.07 &22.59 &22.26  \\
13 &12:43:31.89 &-62:36:14.40 &0.11 &0.27 &0.11 &0.04 &0.02 &23.04 &22.56 &22.23  \\

\hline \multicolumn{7}{c}{G302.48 (1.00e+07 cm$^{-3}$, 70.0 K)}\\\hline
1 &12:47:31.80 &-62:54:0.84 &4.31 &93.89 &24.3 &8.02 &3.77 &24.69 &24.21 &23.88  \\
2 &12:47:31.76 &-62:54:3.32 &6.84 &73.15 &18.93 &6.25 &2.94 &24.21 &23.73 &23.4  \\
3 &12:47:31.97 &-62:54:0.92 &2.74 &27.3 &7.06 &2.33 &1.1 &24.14 &23.65 &23.33  \\
4 &12:47:32.59 &-62:53:57.72 &1.5 &8.32 &2.15 &0.71 &0.33 &23.68 &23.2 &22.87  \\
5 &12:47:31.31 &-62:53:57.00 &0.14 &0.47 &0.12 &0.04 &0.02 &23.28 &22.79 &22.47  \\
6 &12:47:32.01 &-62:53:59.32 &0.2 &0.62 &0.16 &0.05 &0.02 &23.21 &22.72 &22.4  \\
7 &12:47:32.44 &-62:54:1.56 &0.49 &1.47 &0.38 &0.13 &0.06 &23.18 &22.7 &22.37  \\
8 &12:47:32.70 &-62:53:59.96 &0.13 &0.4 &0.1 &0.03 &0.02 &23.15 &22.67 &22.34  \\
9 &12:47:31.35 &-62:53:52.76 &0.17 &0.51 &0.13 &0.04 &0.02 &23.14 &22.66 &22.33  \\
10 &12:47:31.21 &-62:53:54.28 &0.17 &0.46 &0.12 &0.04 &0.02 &23.11 &22.63 &22.3  \\
11 &12:47:31.59 &-62:54:5.64 &0.13 &0.38 &0.1 &0.03 &0.02 &23.1 &22.62 &22.29  \\

\hline \multicolumn{7}{c}{G309.89 (1.00e+07 cm$^{-3}$, 70.0 K)}\\\hline
1 &13:50:35.97 &-61:40:23.72 &7.9 &36.28 &23.91 &7.89 &3.71 &23.88 &23.4 &23.07  \\
2 &13:50:35.71 &-61:40:21.96 &3.0 &14.7 &9.68 &3.2 &1.5 &23.71 &23.23 &22.9  \\
3 &13:50:35.30 &-61:40:21.32 &3.01 &16.85 &11.1 &3.66 &1.72 &23.68 &23.2 &22.87  \\
4 &13:50:35.43 &-61:40:22.84 &3.95 &19.35 &12.75 &4.21 &1.98 &23.58 &23.1 &22.77  \\
5 &13:50:35.69 &-61:40:24.84 &3.05 &12.47 &8.22 &2.71 &1.27 &23.52 &23.04 &22.72  \\
6 &13:50:34.82 &-61:40:19.72 &0.45 &1.48 &0.97 &0.32 &0.15 &23.33 &22.85 &22.52  \\
7 &13:50:37.08 &-61:40:24.52 &0.65 &1.99 &1.31 &0.43 &0.2 &23.31 &22.83 &22.5  \\
8 &13:50:36.02 &-61:40:21.08 &0.51 &1.59 &1.05 &0.35 &0.16 &23.27 &22.79 &22.46  \\
9 &13:50:36.82 &-61:40:24.04 &0.1 &0.2 &0.13 &0.04 &0.02 &22.99 &22.51 &22.18  \\
10 &13:50:35.25 &-61:40:23.48 &0.28 &0.58 &0.38 &0.13 &0.06 &22.97 &22.48 &22.16  \\
11 &13:50:36.52 &-61:40:20.36 &0.13 &0.26 &0.17 &0.06 &0.03 &22.96 &22.48 &22.15  \\
12 &13:50:36.65 &-61:40:18.92 &0.32 &0.64 &0.42 &0.14 &0.07 &22.95 &22.47 &22.14  \\

\hline \multicolumn{7}{c}{G330.28 (1.00e+07 cm$^{-3}$, 86.0 K)}\\\hline
1 &16:3:43.15 &-51:51:46.40 &10.92 &79.9 &53.44 &17.64 &8.29 &24.08 &23.6 &23.27  \\
2 &16:3:43.23 &-51:51:49.68 &3.53 &22.14 &14.8 &4.89 &2.3 &24.04 &23.55 &23.23  \\
3 &16:3:43.10 &-51:51:47.76 &4.02 &26.07 &17.43 &5.75 &2.7 &23.81 &23.33 &23.0  \\
4 &16:3:44.13 &-51:51:45.36 &0.7 &1.52 &1.02 &0.34 &0.16 &23.13 &22.65 &22.32  \\
5 &16:3:43.88 &-51:51:53.12 &0.6 &1.21 &0.81 &0.27 &0.13 &23.12 &22.63 &22.31  \\
6 &16:3:43.82 &-51:51:51.44 &0.29 &0.63 &0.42 &0.14 &0.07 &23.11 &22.63 &22.3  \\
7 &16:3:43.84 &-51:51:46.32 &0.44 &0.8 &0.54 &0.18 &0.08 &22.99 &22.51 &22.18  \\
8 &16:3:44.42 &-51:51:44.64 &0.24 &0.41 &0.28 &0.09 &0.04 &22.99 &22.51 &22.18  \\

\hline \multicolumn{7}{c}{G332.77 (1.00e+07 cm$^{-3}$, 70.0 K)}\\\hline
1 &16:17:31.21 &-50:32:40.02 &0.57 &3.55 &2.57 &0.85 &0.4 &23.61 &23.13 &22.8  \\
2 &16:17:30.96 &-50:32:34.34 &0.19 &0.72 &0.52 &0.17 &0.08 &23.27 &22.79 &22.46  \\

\hline \multicolumn{7}{c}{G336.98 (1.00e+07 cm$^{-3}$, 33.0 K)}\\\hline
1 &16:36:12.41 &-47:37:58.06 &9.57 &243.88 &80.13 &26.45 &12.43 &25.23 &24.75 &24.42  \\
2 &16:36:12.63 &-47:37:57.50 &2.52 &25.73 &8.45 &2.79 &1.31 &24.07 &23.59 &23.26  \\
3 &16:36:12.51 &-47:37:56.70 &2.04 &19.56 &6.43 &2.12 &1.0 &23.94 &23.46 &23.13  \\
4 &16:36:12.09 &-47:37:58.38 &2.2 &12.91 &4.24 &1.4 &0.66 &23.7 &23.22 &22.89  \\

\hline \multicolumn{7}{c}{G337.63 (1.00e+07 cm$^{-3}$, 99.0 K)}\\\hline
1 &16:38:19.11 &-47:4:53.66 &4.26 &78.86 &15.77 &5.2 &2.45 &24.72 &24.23 &23.91  \\
2 &16:38:19.26 &-47:4:57.58 &3.67 &26.27 &5.25 &1.73 &0.81 &24.1 &23.62 &23.29  \\
3 &16:38:19.09 &-47:4:52.70 &1.5 &16.72 &3.34 &1.1 &0.52 &24.07 &23.59 &23.26  \\
4 &16:38:19.09 &-47:4:51.18 &6.62 &53.07 &10.61 &3.5 &1.65 &23.93 &23.45 &23.12  \\
5 &16:38:18.61 &-47:4:52.46 &0.25 &0.66 &0.13 &0.04 &0.02 &23.13 &22.65 &22.32  \\
6 &16:38:18.44 &-47:4:51.98 &0.15 &0.38 &0.08 &0.02 &0.01 &23.09 &22.61 &22.28  \\

\hline \multicolumn{7}{c}{G337.84 (1.00e+07 cm$^{-3}$, 70.0 K)}\\\hline
1 &16:40:26.69 &-47:7:13.42 &2.88 &135.28 &66.43 &21.92 &10.31 &24.99 &24.5 &24.18  \\
2 &16:40:26.95 &-47:7:15.02 &4.1 &73.2 &35.95 &11.86 &5.58 &24.63 &24.15 &23.82  \\
3 &16:40:26.92 &-47:7:16.06 &4.77 &85.62 &42.05 &13.88 &6.52 &24.49 &24.01 &23.68  \\
4 &16:40:26.63 &-47:7:16.14 &3.96 &66.77 &32.79 &10.82 &5.09 &24.19 &23.71 &23.38  \\
5 &16:40:26.64 &-47:7:17.10 &5.08 &64.51 &31.68 &10.45 &4.91 &24.1 &23.62 &23.29  \\
6 &16:40:26.59 &-47:7:14.54 &3.0 &47.28 &23.22 &7.66 &3.6 &24.09 &23.61 &23.29  \\
7 &16:40:26.37 &-47:7:10.38 &0.13 &0.49 &0.24 &0.08 &0.04 &23.3 &22.82 &22.5  \\
8 &16:40:27.46 &-47:7:13.50 &0.12 &0.44 &0.22 &0.07 &0.03 &23.25 &22.77 &22.44  \\
9 &16:40:26.01 &-47:7:9.74 &0.12 &0.41 &0.2 &0.07 &0.03 &23.19 &22.71 &22.38  \\

\hline \multicolumn{7}{c}{G339.11 (1.00e+07 cm$^{-3}$, 50.0 K)}\\\hline
1 &16:42:59.56 &-45:49:44.56 &32.82 &189.31 &104.87 &34.61 &16.27 &23.87 &23.39 &23.07  \\
2 &16:43:0.02 &-45:49:40.32 &11.78 &52.59 &29.13 &9.61 &4.52 &23.58 &23.1 &22.77  \\
3 &16:43:0.79 &-45:49:36.32 &0.13 &0.2 &0.11 &0.04 &0.02 &22.85 &22.37 &22.05  \\
\hline
\hline
\label{tab:clumps}
\end{longtable}
\end{longtab}

\section{Ionised and molecular gas separation}
\label{subsec:molec_removal}

\subsection{\texttt{XCLASS} modelling }
\label{sec:remove_molecules}

\hii{} regions form inside the dense envelopes of MYSOs, often generating or enlarging the region of chemically rich molecular emission, a so-called hot-core.  As such, spectroscopic studies of the youngest \hii{} regions often need to deal with blending of ionised and molecular gas features in the spectrum.  This is the case for many (but not all) of the \hii{} regions in this study. G336.98, amongst the targets with \RRL{29} detections, had the strongest molecular signatures, and is used here to illustrate the separation techniques employed in this study (see Figure \ref{fig:molec_removal}).

As can be seen in the top panel of Figure \ref{fig:molec_removal}, a number of molecular lines overlap in frequency with the much broader RRL emission.  The primary sources of these spectral features are \methacet{} (five lines), CH$_3$OCHO, and in some cases, methanol and ethanol. We only fit \methacet{} in G336.98 because its multiple components allowed us to place greater constraints on the rest velocities and relative intensities of the emission lines. Models with other species that had less constraints on the parameter space failed to converge on repeatable solutions.

Here, we describe the steps involved in fitting the various species, and how we separated the ionised and molecular species for analysis.  The analysis was performed on  smaller subregions within each observed field of view. This subregion corresponds to the region bounded by the continuum core containing the \RRL{29} emission or, in its absence, the brightest core. The exception to this rule was G337.63, where the fourth brightest core contained the \RRL{29} emission.

To ensure our species, fitting had a physical basis rather than producing a (potentially) disjointed set of independent Gaussians that could change dramatically from pixel to pixel, we used the LTE version of  \texttt{XCLASS} (version 1.2) to fit our spectra \citep{XCLASS}. In this process, we found that if the \RRL{29} emission were fit in combination with the molecules, the removal of the molecular species would be relatively straightforward. However, testing showed that in {\it not} fitting the \RRL{29} emission, the resulting spectra (after having removed the poorly fit molecules) had large absorption features in the remaining \RRL{29} spectrum because the intensities of the molecular species had been over estimated to fit the spectra.

RRLs are not included in the databases compatible with \texttt{XCLASS}, which necessitated having to add the $n$ = 29 transition of hydrogen to the local copy of the database.   A new entry was made into the \texttt{sqlite3} database with a transition frequency corresponding to that of \RRL{29}, and dummy values for the components of the required partition function. In this way, \texttt{XCLASS} could simultaneously fit the molecules and \RRL{29} emission, avoiding problems with overestimating the molecular emission at \RRL{29} frequencies.  In no way are the fitted physical properties (i.e. N and T) of the \RRL{29} emission derived from the fitted spectra realistic because the partition function used to derive the physical properties is unrealistic (i.e. populated with zeros).  However, the Gaussian properties of the fit (i.e. FWHM and v\_max) are realistic as they are empirically fit to the data along with the molecular species.  The physical properties derived from the fits for the molecular species are realistic because their partition functions are accurate in the \texttt{XCLASS} database.

The \texttt{XCLASS} routine \texttt{XCLASSMapFit} was used to fit the molecular species mentioned above as well as the \RRL{29} emission. The most robust fitting of a molecule in our \texttt{XCLASS} modelling was that  of \methacet{}, which had five k-ladder transitions in our passband, and therefore also likely gives the best estimate on the physical conditions of the molecular gas surrounding the \hii{} region. This is the case because the properties of the gas surrounding this \hii{} region are good for detecting these transitions of \methacet{}.  Figure \ref{fig:CH3CCH_G336.98} shows the resultant molecular column density, temperature, and velocity offset of the \methacet{} within the modelled region of G336.98 (the same source as shown in Figure \ref{fig:molec_removal}).  Where possible,  the \methacet{} fit properties were used in our analyses (e.g. our Jeans analysis).  The regions with the greatest molecular contributions to the \RRL{29} spectral region were G336.98 and G337.84, and were therefore the targets for which this analysis was most important.

\subsection{Gaussian fitting and spectrum extraction}

\begin{figure}
\includegraphics[width=\columnwidth]{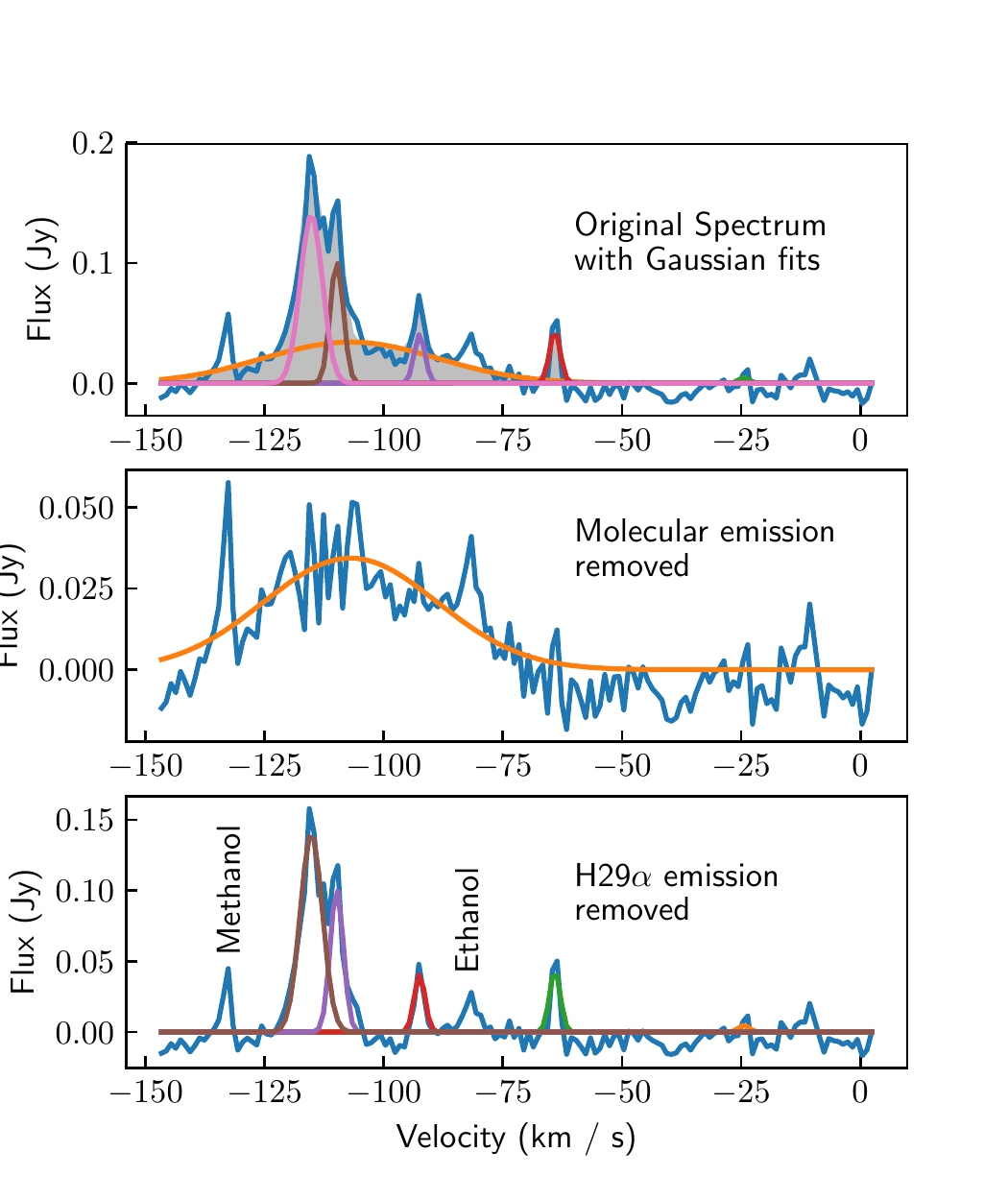}
\caption{\RRL{29} and \methacet{} Gaussian fits to the emission from G336.98. The blue curve shows the overall emission spectrum, the orange curve shows the Gaussian fit to \RRL{29}, and the other colour curves show the Gaussian fits to the \methacet{} components. In the top panel, the blue curve shows the original spectrum; the grey fill indicates the sum of the Gaussian components at each velocity. In the middle panel, the molecular Gaussian components have been removed, and in the bottom panel, the \RRL{29} has been removed, showing how well the individual Gaussian components fit the spectrum. The additional, unfit lines correspond to other species such as methanol and ethanol (at $\sim$ -130 and $\sim$ -80 km s$^{-1}$, respectively), which were not included in the \texttt{XCLASS} fit.}
\label{fig:molec_removal}
\end{figure}

\begin{figure*}
\includegraphics[width=\textwidth]{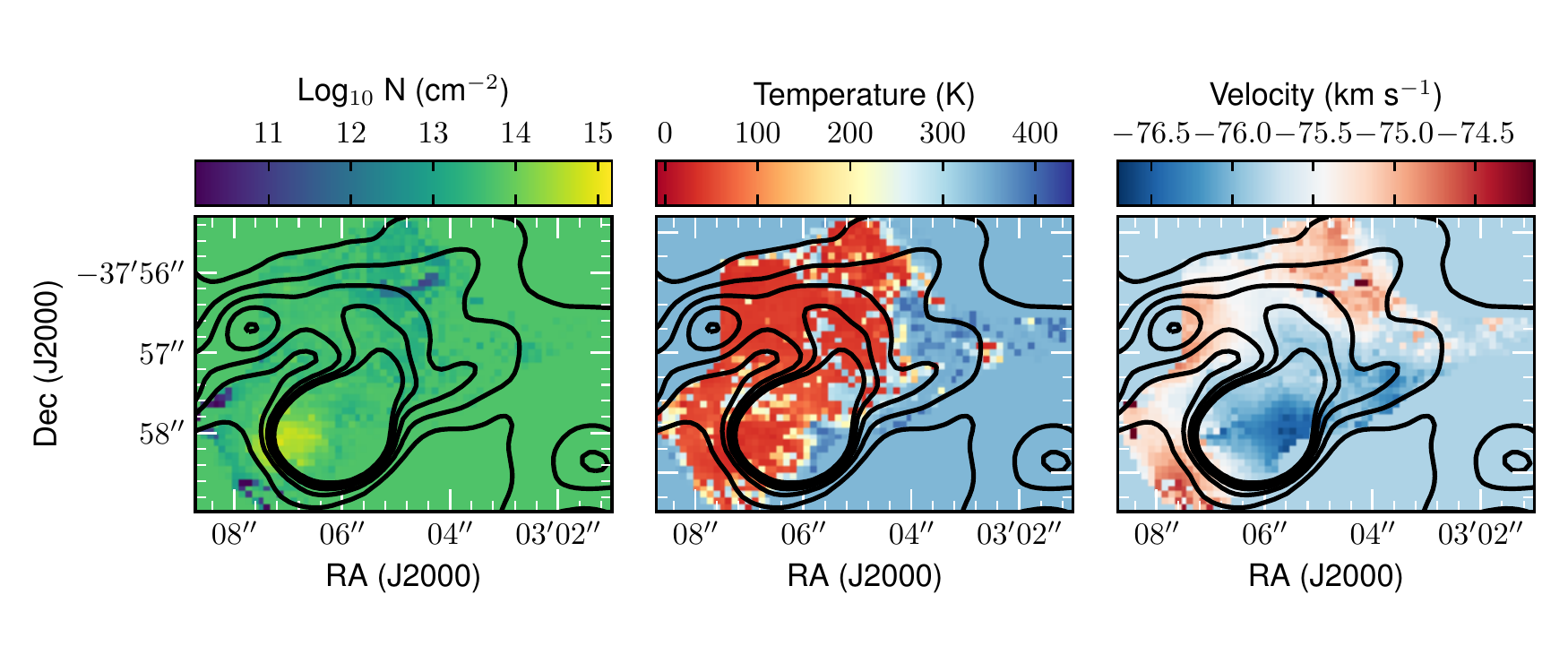}
\caption{Physical properties (e.g. column density, temperature,  and velocity structure from left to right) of the \methacet{} emission derived through \texttt{XCLASS} fitting of the five spectral components shown in Figure \ref{fig:molec_removal} for G336.98. The sharp edge in the temperature and velocity maps indicates the edge of the fitting area rather than a physical jump in parameters. }
\label{fig:CH3CCH_G336.98}
\end{figure*}

The output from \texttt{XCLASS} modelling consisted of a best fit spectrum at each position in the input data cube.  This spectrum is a combination of the emission from each modelled species. As such, a further step of disentangling the \RRL{29} emission from the molecular was required. The Gaussians shown in Figure \ref{fig:molec_removal} represent this subsequent fitting stage after the \texttt{XCLASS} modelling.

These properties were used to constrain subsequent Gaussian fits to the individual components of each species (e.g. the amplitudes, width, and velocities of \methacet{} were applied to each of the five components in our spectral window). This was done for each of the fit species (molecular and ionised) to get the individual Gaussian components that make up the combined spectrum in each pixel.  As shown in the top panel of Figure \ref{fig:molec_removal} (taken from a single pixel in the G336.98 map), the combination of fitting \RRL{29} (orange Gaussian) and \methacet{} (other Gaussians) fits most of the spectral features in this window. The grey filled portion of the spectrum in the top panel shows the profile of the combined Gaussians to highlight that the multi-component fit really is required to account for the ionised and molecular species independently.

With the individual Gaussians computed (using \texttt{pyspeckit} and \texttt{spectralcube}), we can separate the ionised (middle panel) and molecular (bottom panel) components of the emission. The ionised emission spectra (e.g. the original spectrum with the Gaussian fits to the molecular emission removed) were used for the subsequent ionised gas analyses of G336.98 and G337.84.

\end{appendix}

\end{document}